\documentclass[12pt,preprint]{aastex}
\usepackage{epsfig}
                                                                                
\def\vkm{km s$^{-1}$}

\def\degree{$^\circ$}
\def\arcs#1{$#1''$}
\def\arcsa#1#2{$#1^{\prime\prime}_{^\textrm{.}}#2$}

\def\solarmass{$M_\odot$}

\def\mJyb{mJy beam$^{-1}$}

\def\mJybk{mJy beam$^{-1}$ km s$^{-1}$}

\def\cmc{cm$^{-3}$}
\def\cms{cm$^{-2}$}

\def\VLSR{V_\textrm{\scriptsize LSR}}
\def\Vsys{V_\textrm{\scriptsize sys}}
\def\Voff{V_\textrm{\scriptsize off}}

\def\ra#1#2#3#4{#1^\mathrm{h} #2^\mathrm{m} #3^\mathrm{s}_{^\textrm{.}} #4}
\def\dec#1#2#3#4{#1\degr #2\arcmin #3^{\prime\prime}_{^\textrm{.}}#4}

\def\mH2{m_{\textrm{\scriptsize H}_2}}

\def\H2{H$_2$}
\def\N2HP{N$_2$H$^+$}
\def\HCOP{HCO$^+$}
\def\cCO{C$^{18}$O}

\def\NH3{NH$_3$}

\def\SOt{$N_J=8_9-7_8$}

\def\HCOP{HCO$^+$}

\def\putfig#1#2#3{\epsfig{scale=#1,angle=#2,figure=#3}}
\def\putfiga#1#2#3{}
\def\leftblank#1{}

\def\iI{{\it I}}
\def\iQ{{\it Q}}
\def\iU{{\it U}}

\begin{document}

\title{A Pseudodisk Threaded with a Toroidal and Pinched Poloidal Magnetic
Field Morphology in the HH 211 Protostellar System}

\author{Chin-Fei Lee\altaffilmark{1,2}, Woojin Kwon\altaffilmark{3,4},
Kai-Syun Jhan\altaffilmark{2,1}, Naomi Hirano\altaffilmark{1}, Hsiang-Chih Hwang\altaffilmark{5},
Shih-Ping Lai\altaffilmark{6}, Tao-Chung Ching\altaffilmark{7,8}, 
Ramprasad Rao\altaffilmark{1}, and Paul T.P. Ho\altaffilmark{1}}

%, Zhi-Yun Li\altaffilmark{3}, Naomi Hirano\altaffilmark{1}, Hsien Shang\altaffilmark{1},
%Paul T.P.  Ho\altaffilmark{1,4}, and Qizhou
%Zhang\altaffilmark{5}}

\altaffiltext{1}{Academia Sinica Institute of Astronomy and Astrophysics,
P.O. Box 23-141, Taipei 106, Taiwan; cflee@asiaa.sinica.edu.tw}
\altaffiltext{2}{Graduate Institute of Astronomy and Astrophysics, National Taiwan
   University, No.  1, Sec.  4, Roosevelt Road, Taipei 10617, Taiwan}
\altaffiltext{3}{
Korea Astronomy and Space Science Institute (KASI), 776 Daedeokdae-ro, Yuseong-gu, Daejeon 34055, Republic of Korea}
\altaffiltext{4}{
University of Science and Technology, Korea (UST), 217 Gajeong-ro, Yuseong-gu, Daejeon 34113, Republic of Korea}
\altaffiltext{5}{
Department of Physics and Astronomy, Johns Hopkins University, Baltimore, MD 21218, USA}
\altaffiltext{6}{
Institute of Astronomy and Department of Physics, National Tsing Hua University, Hsinchu, Taiwan}
\altaffiltext{7}{
National Astronomical Observatories, Chinese Academy of Sciences, Beijing 100012, Peopleʼs Republic of China}
\altaffiltext{8}{
CAS Key Laboratory of FAST, National Astronomical Observatories, Chinese Academy of Sciences, People’s Republic of China}

%\altaffiltext{3}{Astronomy Department, University of Virginia,
%Charlottesville, VA 22904, USA} \altaffiltext{4}{East Asian Observatory,
%660 N.  A'ohoku Place, University Park, Hilo, HI 96720, USA}
%\altaffiltext{5}{Harvard-Smithsonian Center for Astrophysics, 60 Garden
%Street, Cambridge, MA 02138, USA}

%with Atacama Large Millimeter/submillimeter Array,

\begin{abstract} 

The HH 211 protostellar system is currently the youngest Class 0 system
found with a rotating disk.  We have mapped it at $\sim$ 50 au
(\arcsa{0}{16}) resolution, studying its magnetic field morphology with dust
polarization in continuum at 232 and 358 GHz and its kinematics in \cCO{}
J=2-1 line.  A flattened envelope extending out to $\sim$ 400 au from the
disk is detected in the continuum and \cCO{}, slightly misaligned with the
disk by $\sim$ 8\degree{}.  It is spiraling inwards and expected to
transform into a rotating disk at $\sim$ 20 au, consistent with the disk
radius estimated before.  It appears to have a constant specific angular
momentum and itself can result from an inside-out collapse of an
extended envelope detected before in NH$_3$.  In the flattened envelope, the
polarization is mainly due to the magnetically aligned dust grains,
inferring a highly pinched poloidal field morphology there.  Thus, both the
kinematics and field morphology support that the flattened envelope is a
pseudodisk formed as the infalling gas is guided by the field lines to the
equatorial plane.  Interestingly, a point symmetric polarization
distribution is also seen in the flattened envelope, implying that the
pinched field lines also have a significant toroidal component generated by
the rotation.  No significant loss of angular momentum and thus no clear
magnetic braking are detected in the flattened envelope around the disk
probably because of the large misalignment between the axis of the
rotation and the axis of the magnetic field in the cloud core.

\end{abstract}

%both the kinematics and field morphology support that

%Such a misalignment can also produce the observed misalignment between the
%flattened envelope and the disk.


\section{Introduction}

%in some early models of magnetized core
%collapse,

Magnetic field can produce efficient magnetic braking that can affect star
formation in the early phase.  In particular, magnetic braking can reduce
angular momentum efficiently from a collapsing and rotating envelope,
preventing formation of a Keplerian rotating disk around a protostar, only
allowing a pseudodisk to be formed \citep{Allen2003,Mellon2008}. 
Fortunately, a misalignment between magnetic field axis and rotation axis
\citep{Hennebelle2009,Li2014,Vaisala2019,Hirano2019} and non-ideal MHD
effects, e.g., ambipolar diffusion \citep{Masson2016} and Ohmic
dissipation \citep{Machida2011}, can reduce magnetic braking, enabling
formation of a Keplerian rotating disk.  Recent observations at high angular
resolutions have already detected Keplerian rotating disks as early as in
Class 0 systems, e.g., L1527 \citep{Tobin2012,Sakai2014,Ohashi2014}, HH 212
\citep{Lee2017Disk,Lee2017COM}, and likely HH 211 \citep{Lee2018HH211Disk},
supporting these possibilities.  On the other hand, other observations have
also shown a significant decrease of angular momentum in the HH 111
protostellar system \citep{Lee2010HH111,Lee2016}, supporting a possible
magnetic braking.  However, the detection of a Keplerian rotating disk in
that system \citep{Lee2011} also suggested that the magnetic braking, if
present, does not suppress formation of a disk, but may
constrain the growth of the disk.

Dust polarization observations in millimeter and submillimeter wavelengths
have been used to map the magnetic fields in young protostellar systems in
order to determine the effect of the magnetic field on disk formation in the
beginning of star formation.  Early survey with 4 Class 0 systems showed
that magnetic field axis in the envelopes is more or less aligned with the
outflow axis \citep{Chapman2013}.  However, with a larger sample of 16 Class
0 and I systems, \cite{Hull2013} found that the magnetic field axis in the
envelopes is not necessarily to be aligned with the outflow axis at $\sim$
1000 au scale.  Assuming that the rotation axis in the envelope is aligned
with the outflow axis at $\sim$ 1000 au scale, then the magnetic field axis
in the envelopes is not necessarily to be aligned with the rotation axis of
the envelopes.  Since the amount of magnetic braking depends on the
alignment of the magnetic field axis and the rotation axis in the cloud
core/envelope \citep{Joos2012}, this suggests a different degree of magnetic
braking in different systems, affecting the disk formation and growth in
different levels.  Moreover, detailed mapping toward a few young Class 0
systems, e.g., B335 \citep{Maury2018} and L1448 IRS2 \citep{Kwon2019}, have
detected a pinched poloidal or hour-glass field morphology from a 1000 au
scale down to a 100 au scale, suggesting that magnetic field lines are
affecting the infall motion and are dragged in more along the equatorial
plane than in other directions.  In these Class 0 systems, since the
magnetic field axis in the envelopes is more or less aligned with the
outflow axis, there should be significant magnetic braking.  Since a
presence of a small disk has been hinted at the center in L1448 IRS2
\cite[with a radius $< 100$ au,][]{Tobin2015} and B335 \cite[with a radius
$< 10$ au,] []{Yen2015}, non-ideal MHD effects may be needed in these
systems to reduce the magnetic braking efficiency, allowing a small disk to
be formed at the center, as suggested in the older system HH 111
\citep{Lee2016}.

%NGC1333 IRAS 4A \citep{Girart2006}, It was not detected at 24 \micron{}
%with {\it Spitzer} \citep{Rebull2007} and thus appears to be very similar
%to the PACS Bright Red Sources (PBRS) \citep{Stutz2013}, which seem to be
%among the youngest protostars in Orion.  Therefore, it appears to be among
%the youngest Class 0 protostars in Perseus.

Here, we report the dust polarization observations and envelope kinematics
towards the HH 211 protostellar system, which is one of the youngest Class 0
systems with a rotating disk \citep{Lee2018HH211Disk}, in order to determine
the effect of the magnetic field on disk formation in the earliest phase. 
This system is located at $\sim$ 321$\pm10$ pc away from us
\citep{Ortiz-Leon2018}.  A collimated magnetized jet \citep{Lee2018Bjet} and
a collimated outflow \citep{Gueth1999,Lee2007} are seen extending out from
around the protostar, defining the rotation axis in the system.  This system
seems to have a dynamical collapsing radius of only $\sim$ 1000 au
\citep{Tanner2011}, however, a detection of an infall motion in the envelope
is needed to confirm it.  A marginally resolved disk has been detected at
the center with the Karl G.  Jansky Very Large Array (VLA)
\citep{Segura-Cox2018}.  Observations with Atacama Large
Millimeter/submillimeter Array (ALMA) at higher resolution have resolved it
and found it to be rotating around the protostar, with a small radius of
$\sim$ 20 au \cite[after updated for the new distance,][]{Lee2018HH211Disk}. 
Previous observations with JCMT towards its associated cloud core have
revealed a uniform magnetic field morphology at a 10,000 au scale, with the
field lines roughly north-south oriented and thus with an axis misaligned
with the jet axis by $\sim$ 60\degree{} \citep{Matthews2009}.  At a smaller
scale of a 1000 au, the field lines are still aligned with those in the
larger scale without any clear pinched field morphology \citep{Hull2014}, in
agreement with the source having a small dynamical collapsing radius. 
Further observation within $\sim$ 400 au of the central source at $\sim$ 200
au resolution started to show a hint of a possible pinching and twisting of
magnetic field due to infall and rotation \citep{Lee2014Mag}.  Here we zoom
in to the same inner region to examine the pinching and twisting at $\sim$
50 au ($\sim$ \arcsa{0}{16}) resolution and study the effect of the magnetic
field on disk formation.

% the polarization sensitivity is roughly the same in brightness temperature

%trans 342 0.7 0.51 2 1.15e-3  ==> T =    0.03366 K
% trans 358 0.175 0.111 2 0.07e-3 ==> T =    0.03436 K
% trans 232 0.158 0.104 2 0.018e-3 ==> T =    0.02487 K

\section{Observations}\label{sec:obs}

Polarization observations of HH 211 were executed in Cycle 4 with ALMA 
at $\sim$ 350 GHz in Band 7 and $\sim$ 230 GHz in Band 6.  The project
number was 2016.1.00017.S (PI: Chin-Fei Lee).

%The phase center was $\alpha_{(2000)}=\ra{03}{43}{56}{8040}$,
%$\delta_{(2000)}=\dec{32}{00}{50}{270}$, but the maps here are produced and
%presented with a center at the central source position at
%$\alpha_{(2000)}=\ra{03}{43}{56}{8054}$,
%$\delta_{(2000)}=\dec{32}{00}{50}{189}$ \citep{Lee2018HH211Disk}.

% In Technical Handbook
% C40-7 
% HH 211 Elev = 30degree
% L5 = 5th percentile: 379m ==> 189 m  ==> 158 m
% L80 = 80th percentile: 1748.2 ==> 874 m ==> 733 m
% Maximum baseline: 3696.9m ==> 1848 m ==> observed projected uvdist 1550 m
% theta_MRS = 0.983*lambda/L5 = 1.1"
% theta_res ~ 0.574*lambda/L80 = 0.14"

%(Low dec source, elevation is 29-35 degree, with a mean of 32degree.

%Observing with an elevation of about $\sim$ 32\degree{}, the maximum
%recoverable scale is $\sim$ \arcsa{1}{2}.

%Since the jet consists of a chain of knots and subknots, the detection of
%polarized emission towards the subknots, which have a size of $\sim$
%\arcsa{0}{2}, will not be affected.  However, since the jet seems to have a
%smooth structure greater than the maximum recoverable scale, the polarized
%emission there, if exists, could be partially resolved out.

\subsection{Band 7}

% One pointing with a primary beam (a field of view) of $\sim$ \arcs{17} was
%observed in order to map a circular region within $\sim$ \arcs{8} of the
%central source.

HH 211 was observed in Band 7 with two executions on 2016 October 10,
with 43 antennas in C40-7 configuration and a total time of $\sim$ 77
minutes.  It was observed with one single pointing towards the central
source.  The primary beam had a size of $\sim$ \arcs{17}.  The projected
baselines were $\sim$ 20-1570 m.  The maximum recoverable scale was $\sim$
\arcsa{1}{1}.  There were 5 spectral windows in the correlator setup (see
Table \ref{tab:corrb7}).  Here, we only report the results in continuum for
the envelope and disk.

We used the CASA package (versions 4.7) to calibrate the data.  Quasars
J0238+1636, J0237+2848 ($\sim$ 0.833 Jy), J0336+3218 ($\sim$ 0.517 Jy), and
J0334$-$4008 ($\sim$ 0.415 Jy) were used to calibrate the flux, the
passband, the gain, and the polarization, respectively.  In order to improve
the map fidelity, we also did a phase-only self-calibration on the data with
the continuum intensity (Stokes {\it I}) map.  A visibility weighting with a
robust factor of 0.5 was used to make the continuum maps (including Stokes
\iI{}, \iQ{}, and \iU{} parameters).  The resulting synthesized beam
(resolution) has a size of $\sim$ \arcsa{0}{18}$\times$ \arcsa{0}{11}. 
Since the envelope and disk were detected only within $\sim$ \arcsa{1}{5} of
the central source, we did not perform any primary beam correction on the
maps.  The Stokes \iI{} map has a noise level of $\sim$ 0.16 \mJyb{}.  The
Stokes \iQ{} and \iU{} maps have a noise level of $\sim$ 0.07 \mJyb{}.

\subsection{Band 6}

HH 211 was also observed in Band 6 with two executions on 2017 August
08, with 46 antennas in C40-8 configuration and a total time of $\sim$ 63
minutes.  It was observed with one single pointing towards the central
source.  The primary beam had a size of $\sim$ \arcs{28}.  The projected
baselines were $\sim$ 20-3270 m.  The maximum recoverable scale was $\sim$
\arcsa{1}{8}.  There were 6 spectral windows in the correlator setup (see
Table \ref{tab:corrb6}).  Here, we only report the results in continuum and
\cCO{} for the envelope and disk.

We used the CASA package (versions 4.7) to calibrate the data.  Quasars
J0238+1636, J0237+2848 ($\sim$ 1.219 Jy), J0336+3218 ($\sim$ 0.823 Jy), and
J0522$-$3627 ($\sim$ 6.05 Jy) were used to calibrate the flux, the passband,
the gain, and the polarization, respectively.  In order to improve the map
fidelity, we also did a phase-only self-calibration on the data with the
continuum intensity (Stokes {\it I}) map.  A visibility weighting with a
robust factor of 0.5 was used to make the continuum maps (including Stokes
\iI{}, \iQ{}, and \iU{} parameters) and the \cCO{} channel maps.  The
resulting synthesized beam (resolution) has a size of $\sim$
\arcsa{0}{16}$\times$ \arcsa{0}{10} in continuum and $\sim$
\arcsa{0}{17}$\times$ \arcsa{0}{11} in \cCO{}.  Since the envelope and disk
were detected only within $\sim$ \arcsa{1}{5} of the central source, we did
not perform any primary beam correction on the maps.  For the continuum, the
Stokes \iI{} map has a noise level of $\sim$ 0.030 \mJyb{}, while the Stokes
\iQ{} and \iU{} maps have a noise level of $\sim$ 0.018 \mJyb{}.  The
velocity resolution and noise level in the \cCO{} channel maps are 0.2
\vkm{} per channel and $\sim$ 1.8 \mJyb{}, respectively.

\subsection{Polarization Measurements}

Polarization intensity is bias-corrected with $P_i =
\sqrt{Q^2+U^2-\sigma_p^2}$, where $\sigma_p$ is the noise in the unbiased
polarization intensity.  Thus, we consider it a detection if $P_i \geq 2.5
\sigma_p$.  Polarization degree (fraction) is defined as $P_d=P_i/I$. 
Polarization orientations are given by the $E$ vectors.  Based on the ALMA
Cycle 4 Technical Handbook, we can achieve an accuracy better than 0.3\%
(3$\sigma$) in polarization detection within the inner 1/3 of the
primary beam.

%and this determines the largest acceptable angular size of sources which
%can be observed in full polarization.  }

\section{Results}

In this system, the jet has a position angle (P.A.) of $\sim$ 116.6\degree{}
\citep{Lee2009HH211} and is inclined at $\sim$ 9\degree{} to the plane of
the sky, with its southeastern side tilted slightly away from it
\citep{Jhan2016}.  The outflow has roughly the same axis as the jet
\citep{Gueth1999,Lee2007}.  The disk has a radius of $\sim$ 20 au and a
major axis with a P.A.  of $\sim$ 27.6\degree{} \citep{Lee2018HH211Disk}. 
Thus, the disk is almost perpendicular to the jet axis and should be nearly
edge-on with an inclination of $\sim$ 81\degree{} to the plane of the sky,
with its nearside tilted slightly towards the northwest.  The central
protostar is located at $\alpha_{(2000)}=\ra{3}{43}{56}{8054}$ and
$\delta_{(2000)}=\dec{32}{00}{50}{189}$, as previously found at $\sim$
\arcsa{0}{03} resolution \citep{Lee2018HH211Disk}.

%It has a Gaussian deconvolved radius of $\sim$ 19 au (i.e., \arcsa{0}{06}).

%roughly aligned with the disk major axis and thus almost exactly
%perpendicular to the jet axis, tracing the dust emission in the envelope
%and disk.

\subsection{Continuum Intensity Maps}

\begin{figure} [!hbp]
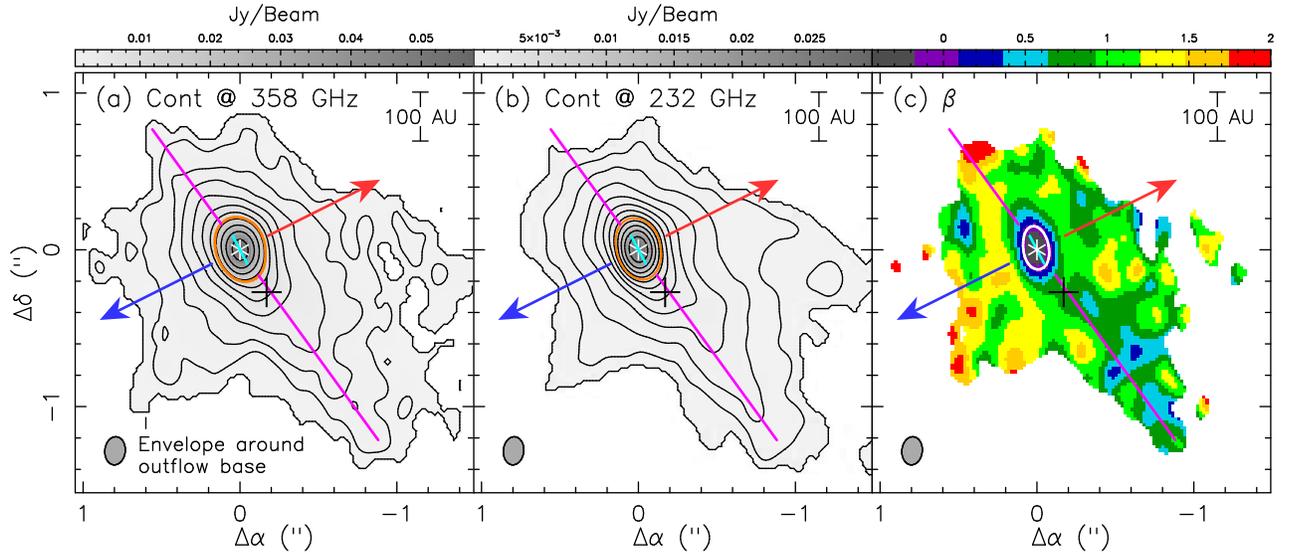

\centering
\putfig{0.7}{270}{f1.eps} %{cont_beta.eps}
\figcaption[]
{Continuum intensity maps at 358 and 232 GHz at a resolution of
\arcsa{0}{18}$\times$\arcsa{0}{13} and the resulting $\beta$ map.  White
asterisk marks the central source position.  Black cross marks the secondary
source previously detected in \citet{Lee2009HH211}.  The magenta lines show
the major axis of the envelope.  The cyan lines centered at the source
show the major axis of the disk.  The blue and red arrows show the
blueshifted side and the redshifted side of the jet axis in the inner part. 
The contour levels are $5\cdot1.5^{n-1}\sigma$, with $n=1,2,3..$, where
$\sigma=0.16$ \mJyb{} (0.0652 K) in (a) and $\sigma=0.047$ \mJyb{} (0.0484
K) in (b).  In panels (a) and (b), the orange contours mark the
boundaries where the brightness temperature is 5 K.  In panel (c), the
outermost region in the $\beta$ map has been masked out using the second
lowest contour in the 358 GHz map for reliable $\beta$ values.  Also, the
white contour around the central source marks the boundary where $\beta =
0$.
\label{fig:contbeta}}
\end{figure}

Figure \ref{fig:contbeta} shows the continuum intensity maps toward the
center at 358 and 232 GHz.  The maps were made with the same $uv$ range of
$\sim$ 24 to 1850 $k\lambda$ and then convolved to the same angular
resolution of \arcsa{0}{18}$\times$\arcsa{0}{13} for comparison and dust
opacity index calculation below.  The continuum emission is detected in
roughly the same regions with a similar morphology at the two frequencies. 
It is highly peaked at the center due to the bright compact disk
\citep{Lee2018HH211Disk}.  It extends $\sim$ \arcs{1} to the northeast and
$\sim$ \arcsa{1}{5} to the southwest from the central source, tracing a
flattened envelope with a major axis at a P.A.  $\sim$ 36\degree{} (as
indicated by the magenta lines), slightly rotated counterclockwise by $\sim$
8\degree{} from the disk major axis (as indicated by the cyan lines).  The
envelope has a brightness temperature of less than 5 K, except for the
central region inside the orange contours shown in Figures
\ref{fig:contbeta}a and \ref{fig:contbeta}b.  The flattened envelope is
asymmetric, extending more to the southwest.  It could be due to a secondary
source (marked with a cross) previously detected at $\sim$ \arcsa{0}{3} to
the southwest in the flattened envelope with the Submillimeter Array (SMA)
at a slightly lower resolution \citep{Lee2009HH211}.  However, this
secondary source is not confirmed here and thus further observation is
needed to check the existence of it.  The continuum emission also
traces the envelope material around the outflow base (as indicated in Figure
\ref{fig:contbeta}a) previously detected in CO \citep{Gueth1999,Lee2007},
forming a small U-shaped shell-like structure opening to the southeast
around the jet axis. A possible similar shell-like structure is also seen
extending to the northwest around the jet axis, especially at 232 GHz, but
not as clear as that seen extending to the southeast.

%As suggested in \citet{Lee2018HH211Disk}, it could be an artifact due to
%limited $uv$-coverages of the SMA when using the super-uniform weighting on
%the visibilities.

\subsection{Spectral Index and Dust Opacity Index}\label{sec:index}

The morphology of the continuum emission is similar at the two frequencies,
allowing us to derive the spectral index of the continuum with the following
formula:

\begin{equation} 
\alpha = \frac{\log\frac{I_{\nu_2}}{I_{\nu_1}}}{\log\frac{\nu_2}{\nu_1}}
\end{equation}

\noindent where $I$ is the specific intensity, $\nu_2 = 358$ GHz, and $\nu_1
= 232$ GHz.  Then we estimate the dust opacity index with $\beta=\alpha-2$,
assuming Rayleigh-Jean limits for the two frequencies.  As shown in Figure
\ref{fig:contbeta}c, $\beta \lesssim 0$ within $\sim$ \arcsa{0}{15} of the
center (as marked by the white contour) because the disk there is optically
thick \citep{Lee2018HH211Disk} and hence the dust opacity index can
not be derived properly.  In the envelope outside the disk region, we have
$\beta$ $\gtrsim$ 1 mostly.  Similar $\beta$ values have also been found
before in the envelopes of the Class 0 sources \citep{Kwon2009} and the
Class I sources \citep{Agurto2019}, suggestive of some degree of grain
growth in the envelopes.

% with a maximum grain size $\lesssim 100$ \micron{}.}

%Notice that since most of the envelope can have a temperature as low as
%20$-$50 K, the Rayleigh-Jean limits become inaccurate and the $\beta$ value
%derived here can be underestimated by $\sim$ 0.35$-$0.15.  Thus, the actual
%$\beta$ $\gtrsim$ 1.0 for most of the envelope.

% Since the envelope is cold and could have a temperature of 20 K, the
% Rayleigh-Jean limits become inaccurate and the $\beta$ value derived here
% could be underestimated by $\sim$ 0.4.  with a temperature of about 20 K
% and thus the beta value here could be $\sim$ 1.0.

\subsection{Polarization Detections}

Polarized dust emission is detected at both frequencies, with more
detections at 232 GHz because of a higher sensitivity (i.e., a lower noise
level), arising roughly from the same regions, as shown in Figure
\ref{fig:contpol}.  It is clearly detected towards the optically thick
region within $\sim$ \arcsa{0}{15} of the center, where the disk is located,
and extends $\sim$ \arcsa{0}{7} to the northeastern envelope.  There are
also patchy detections in the southwestern envelope and some other regions. 
The polarization degree is $\sim$ 2\% at the center where the disk is
located and it increases to more than 10\% going away from the disk to the
outer edges of the envelope.  Such a rapid increase in polarization degree
from the disk to the envelope has also been seen in a survey of 10
protostellar systems \citep{Cox2018}, and can be used to discriminate the
polarization mechanism between the disk and the envelope.  Comparing Figure
\ref{fig:contpol}c with Figure \ref{fig:contpol}d, we find that the
polarization orientations at the center are roughly the same at the two
frequencies, oriented roughly in the east-west direction, with those at 358
GHz being slightly more aligned with the minor axis of the disk than those
at 232 GHz by $\sim$ 8\degree{}.  Going away to the northeastern
envelope from the center, the polarization orientations rotate
counter-clockwise to be more perpendicular to the major axis of the
flattened envelope.  In the southwestern envelope, polarizations at 358 GHz
are detected along the midplane, with their orientations roughly parallel to
the major axis of the envelope.  On the other hand, the polarizations at 232
GHz are slightly away from the midplane, with their polarizations roughly
east-west orientated.  Interestingly, there is a clear point symmetry in the
polarization distribution at 232 GHz in the envelope along the equatorial
plane, with fewer detections on the east side of the northeastern envelope
and fewer detections on the west side of the southwestern envelope.  In
addition, there seems to be a polarization hole at both frequencies centered
at the tentative secondary source.  Moreover, almost no polarization is
detected along the outflow axis more than $\sim$ \arcsa{0}{1} above and
below the central source.

% (see Figure \ref{fig:contpol}).

%That no polarized dust emission is detected elsewhere could be
%due to insufficient sensitivity and insensitivity of our observations to a
%size scale larger than $\sim$ \arcs{1}.

\subsection{Envelope Kinematics in \cCO{}} \label{sec:env}

%calc "1.127*0.085*235/1000*1.127*1.5**2" ==> 0.057 Ms   (in Lee et al. 2018 with d=235pc)
%calc "1.127*0.085*321/1000*1.127*1.5**2" ==> 0.078 Ms   Adjusted to d=321

In our observations at 232 GHz, \cCO{} J=2-1 line was also observed
simultaneously, allowing us to determine the kinematics of the envelope and
how it affects the field morphology to be inferred from the polarizations. 
The systemic velocity was previously found to be $\Vsys \sim 9.14$ \vkm{} in
\citet{Tanner2011}.  It is refined slightly here to be $\Vsys \sim 9.10$ \vkm{}
LSR in order to fit the \cCO{} kinematics, as discussed later.  In order to
simplify our presentations, we define an offset velocity
$\Voff=\VLSR-\Vsys$.

The integrated intensity, blueshifted, and redshifted maps of \cCO{} are
shown on top of the 232 GHz continuum map in Figure \ref{fig:ccoenv}.  As
seen in Figure \ref{fig:ccoenv}a, the integrated intensity map shows that
the \cCO{} emission, although less extended to the southwest probably
because of an insufficient sensitivity, forms a similar structure to the
continuum and thus traces the envelope reasonably well.  Almost no emission
is detected towards the central continuum peak because of a self-absorption
against the bright and optically thick disk at the center, as previously
seen in SO \citep{Lee2018HH211Disk}.  As seen in Figures \ref{fig:ccoenv}b
and \ref{fig:ccoenv}c, the blueshifted emission and redshifted emission are
mostly on the opposite sides of the flattened envelope, because of a
rotation motion.  The overlaps of the blueshifted and redshifted emission in
the envelope is due to the presence of an infall motion.

Position-velocity (PV) diagram along the major axis of the envelope is
presented in Figure \ref{fig:pvenv}a to show the brightness temperature
distribution as well as the infall and rotation motion in the envelope.  As
can be seen from the diagram, the envelope has a brightness temperature of
$\sim$ 20$-$30 K in the inner part and $\sim$ 10 K in the outer part,
indicating that the envelope is warm.  The blueshifted peak is higher than
the redshifted peak and an absorption dip is seen on the redshifted side at
$\sim$ 0.6 \vkm{} due to the self-absorption against the bright disk at the
center, as expected for an infalling envelope \citep{Lee2014HH212}.  The PV
structure can be roughly described as two triangular structures, with one on
the blueshifted side and the other on the redshifted side, with their tips
pointing away from each other, also expected for an infalling envelope
\citep{Ohashi1997}.  In addition, since the triangular structures are
shifted slightly to the southwest on the blueshifted side and to the
northeast on the redshifted side, the envelope also has a rotation motion.

%calc "0.8*321*0.19*321/280" = 56 au km/s  from Lee et al. 2009 HCO+ Figure 4a

This type of PV structure has been seen previously in older systems, e.g.,
HH 212 \citep{Lee2014HH212,Lee2017COM} and L1527 \citep{Sakai2014}.  It has
been well fitted by a simple ballistic infalling-rotating envelope model
\citep{Sakai2014,Lee2017COM}, in which the material is assumed to spiral
inwards with constant specific angular momentum and total energy.
  In this model, only two parameters are needed, one is the specific angular
momentum of the envelope, $l$, and the other is the mass of the central
protostar, $M_\ast$ \cite[for the equations, please
see][]{Sakai2014,Lee2017COM}.  The mass of the central protostar was
previously estimated to be $M_\ast\sim$ 0.08 \solarmass{} \cite[after
updated with the new distance,][]{Lee2018HH211Disk}, assuming a Keplerian
rotation for the disk.  With this mass, the outer boundaries of the PV
structure can be fitted reasonably well with $l \sim 55\pm15$ au
\vkm{}, which is also well consistent with previous measurement of angular
momentum in \HCOP{} \cite[see the solid curve in Figure 4a
in][]{Lee2009HH211}.  Notice that our model fitting is based mainly on the
northeastern envelope because the southwestern envelope could have been
affected by a tentative secondary source.  In addition, the systemic
velocity is refined slightly to be $\sim$ 9.10 \vkm{} LSR.  Our fitting
result not only allows us to derive the infall and rotation velocities in
the envelope, but also confirms the previously estimated mass of the central
protostar.

{The infall and rotation velocities of} this model are shown in Figure
\ref{fig:pvenv}b.  As can been seen, the centrifugal radius, where the
infall velocity is the same as the rotation velocity, is
$r_c=\frac{l^2}{GM_\ast}\sim$ \arcsa{0}{13} or 42 au.  Thus, the infall
velocity is larger than the rotation velocity outside this radius, but
smaller than the rotation velocity inner to this radius.  The centrifugal
barrier, where the infall velocity becomes zero and a rotating disk is
expected to form, has a radius of $r_b=\frac{l^2}{2GM_\ast}\sim$
\arcsa{0}{067} or 21 au, also well consistent with the radius of the dusty
disk \citep{Lee2018HH211Disk}.  Thus, the model result also supports that
the infalling and rotating flattened envelope has transformed into a
rotating disk at $\sim$ 20 au.

%The region in between $r_c$ and $r_b$ can be considered as a transition

%Magnetic field is frozen in to the collapsing material and the collapse is
%guided by the field lines. 

\subsection{Magnetic Field Morphology}

In the envelope of this very young system, since the dust opacity index is
mostly greater than 1 (see Figure \ref{fig:contbeta}c), the dust grains are
likely to be too small to have significant dust self-scattering in our
polarization observations, especially at the lower frequency of 232 GHz. 
Thus, the polarizations are likely due to magnetically aligned dust grains,
as found in the envelopes in other Class 0 sources
\citep{Cox2018,Maury2018,Kwon2019}.  In addition, with a brightness
temperature of less than 5 K and thus much less than that seen in \cCO{},
the continuum emission is optically thin in the envelope.  Hence the
inferred magnetic field orientations are perpendicular to the polarization
orientations, as shown in Figure \ref{fig:Bfield}.  In this figure, the
magnetic field orientations inferred from the polarization orientations at
both 358 GHz and 232 GHz are plotted together to show a more complete
picture of the field morphology.  For comparison, the magnetic field
orientations previously detected with SMA at 342 GHz at more than
2.5$\sigma$ and $\sim$ \arcsa{0}{6} resolution \citep{Lee2014Mag} are also
included as orange lines.  From the figure, it is clear that the previously
detected magnetic fields with SMA are mostly located near the outer edges of
the envelope and are roughly aligned with those detected here at 232 and 358
GHz with ALMA.

%the inferred magnetic field has an
%orientation neither parallel nor perpendicular to the disk major axis
%\cite[see Figure \ref{fig:Bfield}b for the zoom-in plotted over the
%higher-resolution disk map adopted from][with the cyan line showing the disk
%major axis]{Lee2018HH211Disk}.  This inferred field orientation is very
%uncertain because of the following reasons.  

%In addition, since the disk is optically thick, the inferred magnetic field
%orientation, even if the dust polarization is really due to magnetically
%aligned grains, should be parallel to the polarization orientation instead
%of perpendicular \citep{Lee2018BDisk}.

\begin{figure} [!hbp]
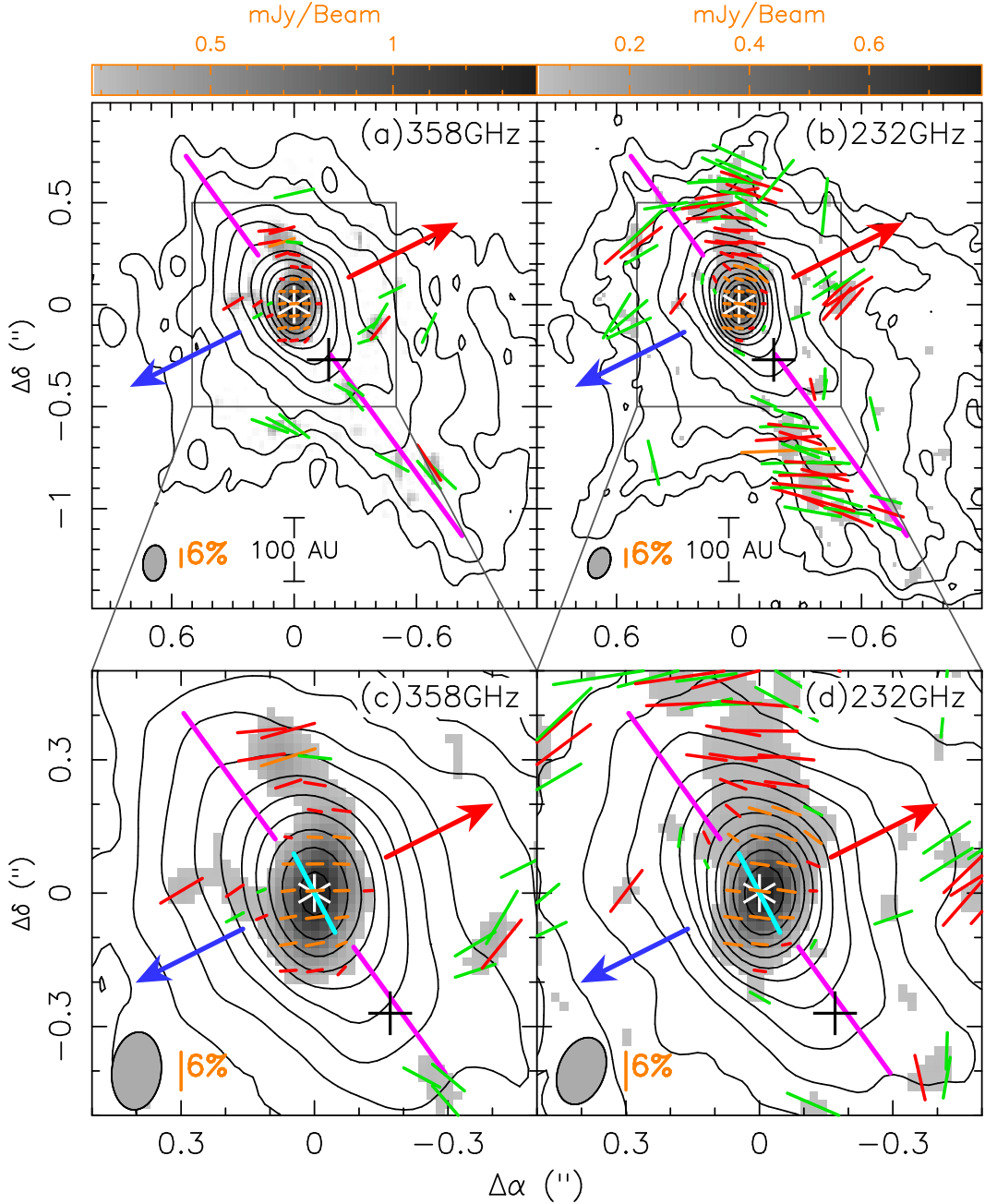

\centering
\putfig{0.92}{270}{f2.eps} %{cont_pol.eps}
\figcaption[]
{Polarization detections toward the envelope and disk in the HH 211 system
at 358 GHz (at a resolution of $\sim$ \arcsa{0}{18}$\times$\arcsa{0}{11})
and 232 GHz (at a resolution of $\sim$ \arcsa{0}{16}$\times$\arcsa{0}{10}). 
Panels (c) and (d) are the zoom-ins of panels (a) and (b).  The asterisk,
the cross, the magenta lines,  the red and blue arrows, and the cyan lines
have the same meanings as in Figure \ref{fig:contbeta}.
Contours show the continuum intensity maps.  The
contour levels are $5\cdot1.5^{n-1}\sigma$, with $n=1,2,3..$, where
$\sigma=0.16$ \mJyb{} in (a) and $\sigma=0.03$ \mJyb{} in (b). 
Grayscale image shows the polarized intensity with more than 2.5$\sigma$
detection. The line segments show the polarization orientations, with green
for 2.5$-$3$\sigma_p$ detections, red for 3$-$5$\sigma_p$ detections, and
orange for more than 5$\sigma_p$ detections, where the noise in the
polarization $\sigma_p=0.07$ \mJyb{} in (a) and $\sigma_p=0.018$ \mJyb{} in
(b).
\label{fig:contpol}}
\end{figure}

On the other hand, in the innermost region within $\sim$ \arcsa{0}{1} of the
center, where the optically thick disk is located, the polarization can
be significantly affected by dust self-scattering, as seen in Class 0/I
disks \citep{Sadavoy2018,Lee2018BDisk,Cox2018} and Class I/II disks
\citep{Stephens2017,Kataoka2017,Hull2018,Bacciotti2018}, because the grains
might have grown to submillimeter size in the disk.  In particular, the
observed polarization orientations are more aligned with the disk minor axis
at higher frequency (358 GHz) than lower frequency (232 GHz) (comparing
Figure \ref{fig:contpol}c and Figure \ref{fig:contpol}d), as expected if
dust self-scattering did contribute partly to the dust polarization.  In
addition, since the innermost region is not spatially resolved, the
polarization orientation is a combination of those in the optically thick
disk and those in the optically thin innermost envelope.  As a result,
observations at higher resolution and longer wavelengths are needed to
determine the origin of the dust polarization and infer the magnetic field
there properly.

The envelope is the region at more than $\sim$ \arcsa{0}{1} away from the
center.  The polarization orientations at 358 GHz are also slightly rotated
counter-clockwise with respect to those at 232 GHz (comparing Figure
\ref{fig:contpol}c and Figure \ref{fig:contpol}d), and thus could be
slightly affected by dust self-scattering.  Thus, we focus more on the
detections at 232 GHz because it is less affected by dust self-scattering,
if any.  In the northeastern envelope, a pinched field morphology (as
indicated by the white curves) is seen, with the field lines converging from
the outer edge of the envelope down to the northeastern edge of the disk. 
Deeper polarization observations are needed to consolidate this pinched
field morphology by detecting more polarization on the east side of the
envelope.  In the southwestern envelope, a pinched field morphology (as
indicated by the white curves) is also seen extending from the outer edge of
the envelope towards the disk down to within $\sim$ \arcsa{0}{4} of the
center.  Again, deeper polarization observations are needed to consolidate
this pinched field morphology by detecting more polarization on the west
side of the envelope.  In addition, since a polarization hole is seen
further in around the tentative secondary source, deeper observations are
needed to check if the pinched field morphology extends down to the disk. 
The field lines above and below the envelope midplane near the jet axis,
although patchy, are roughly perpendicular to the jet axis and thus may
trace toroidal fields in the envelope wrapped around the jet axis, as
discussed later.  The field lines at $\sim$ \arcsa{0}{5} towards the
southeast (as marked ``P" in Figure \ref{fig:Bfield}a) are located in the
envelope around the outflow base, likely tracing the poloidal fields there,
as discussed later.

\section{Discussion}

\subsection{Inside-Out Collapse}\label{sec:collapse}

%$\sim$ $\frac{4}{3}$ times the radius of the keplerian disk
%\citep{Lee2017disk}.

% and can be roughly fitted with the same relationship as that in the NH$_3$
%cores \citep{Goodman1993}, which is $l\propto r^{1.6}$ (the sold curve).

In HH 211, the flattened envelope detected here is deeply inside a rotating
extended envelope, which has a size up to $\sim$ 10$^4$ au detected before
in NH$_3$ \citep{Wiseman2001,Tanner2011}, as shown in Figure \ref{fig:nh3}. 
Interestingly, the extended envelope is flattened and warped, with the outer
part (especially in the northeast) close to perpendicular to the large-scale
magnetic field axis (as shown with the line segments) and the inner part
close to perpendicular to the jet axis.  Figure \ref{fig:rcollapse} presents
the specific angular momentum distribution in the \cCO{} flattened envelope
(horizontal bar) in comparison to that in the NH$_3$ extended envelope (data
points with error bars) reported in \citet{Tanner2011}.  As discussed
earlier in Section \ref{sec:env}, the \cCO{} flattened envelope is found to
have a roughly constant specific angular momentum of $\sim 55\pm15$ au
\vkm{} up to a radius of $\sim$ 400 au, with its PV structure well fitted
with a ballistic infalling-rotating model.  As can be seen, the extended
envelope has a different distribution of specific angular momentum, with its
specific angular momentum decreasing from the outer part to the inner part,
down to the value roughly matching that in the flattened envelope.  Since
the extended envelope is gravitational stable \citep{Tanner2011} and the
flattened envelope is collapsing towards the center, this indicates that the
inner part of the extended envelope has transformed into the collapsing
flattened envelope, confirming the previous study \citep{Tanner2011}.

%Here we assume the gravity dominates the magnetic energy
%and rotational energy.

Inside-out collapsing model has been used to explain the collapse of a cloud
core \citep{Shu1977}.  It has an analytic self-similar solution and thus can
be used to roughly estimate the collapsing age $t_c$, collapsing radius
$r_c$, and other parameters.  This model starts with a singular isothermal
sphere.  An expansion wave is initiated at the center and propagates
outwards at the isothermal sound speed $a$ and then the material within the
expansion wave collapses inside-out towards the center.  The collapsing
radius is the radius where the expansion wave has propagated to and where
the material starts to fall in.  The collapsing age is the time elapse since
the wave is initiated at the center when the mass of the protostar is zero,
and is thus the same as the age of the protostar.  This model has a
self-similar solution for the infall motion \cite[see Equation 8 and Table 2
in][]{Shu1977}, allowing us to compute the radial position for each
infalling material (and its specific angular momentum) at given $t_c$.

Here, the NH$_3$ extended envelope can be identified as the cloud core
and the \cCO{} flattened envelope can be identified as the collapsing
envelope within the expansion wave.  The NH$_3$ extended envelope has a mean
kinetic temperature of $\sim$ 15 K \citep{Tanner2011}, and thus an
isothermal sound speed $a \sim$ 0.23 \vkm{}.  In this extended envelope, the
distribution of specific angular momentum can be roughly fitted with

\begin{equation}
l_e \sim 72 \Big(\frac{r}{1000 \;\textrm{au}}\Big)^{1.55} \;\textrm{au \vkm{}}
\end{equation}

\noindent as shown by the solid curve in Figure \ref{fig:rcollapse}.  The
power-law index here is almost the same as that found in the survey of
NH$_3$ cloud cores \citep{Goodman1993}.  With this extended envelope as an
input to the inside-out collapsing model, we have calculated the
distribution of specific angular momentum in the collapsing envelope within
the expansion wave at 3 collapsing ages, as shown by the 3 dotted curves in
Figure \ref{fig:rcollapse}.  As can be seen, the model at an older age
produces a larger specific angular momentum in the collapsing envelope. 
Interestingly, for a given age, the specific angular momentum in the inner
part of the collapsing envelope appears almost constant.  This is because
the material in this part of the collapsing envelope comes from a thin shell
(with a small range of radii) in the original extended envelope and thus has
roughly the same specific angular momentum.  For example, at $t_c \sim
35000$ yrs, the thin shell that forms this part of the collapsing envelope
has a small range of radii around 830 au, as marked by the small thick line
in the figure.  This thin shell of the original extended envelope falls in
and is spread across the inner part of the collapsing envelope because the
inner part of the shell falls earlier and thus faster than the outer part
\cite[see Figure 3b in][]{Shu1977}.  Since the inner radius of this thin
shell is given by $r_x = 0.4875\,a\,t_c$, the specific angular momentum in
the inner part of the collapsing envelope is $\sim l_e(r_x) \propto
t_c^{1.55}$, and thus increasing with the collapsing age.

%Notice that this collapsing age is much longer than the dynamical age of
%the jet and outflow, which were found to be $\sim$ 1000 yrs
%\citep{Gueth1999}.

In this inside-out collapsing model, the \cCO{} flattened envelope can be
regarded as the inner part of the collapsing envelope where the specific
angular momentum is roughly constant.  As can be seen, the collapsing model
here at $t_c \sim 35000$ yr fits the observed distribution of specific
angular momentum in the \cCO{} flattened envelope reasonably well.  In this
case, the expansion wave has propagated to $r_c \sim$ 1700 au (as marked by
the circle in Figure \ref{fig:nh3}) and the flattened envelope comes from a
thin shell originally located at $r\sim r_x = 830$ au.  The material with $r
\leq r_x$ in the original extended envelope has already accreted to the
central protostar.  The accretion rate predicted in the inside-out model is
$0.975\,a^3/G \sim 2.81\times 10^{-6}$ \solarmass{} yr$^{-1}$.  The observed
mean accretion rate to form the central protostar is $\dot{M} \sim
M_\ast/t_c \sim$ 2.25$\times10^{-6}$ \solarmass{} yr$^{-1}$, which is $\sim$
80\% of that predicted, probably because part of the material goes into the
jet and outflow, carrying away excess angular momentum.  With this observed
accretion rate and a reasonable protostellar radius of $R_\ast \sim 2
R_\odot$, the accretion luminosity would be $\sim G M_\ast \dot{M}/R_\ast
\sim 2.73$ $L_\odot$, about 60\% of the bolometric luminosity observed
before \citep{Froebrich2003}.  Notice that since the inside-out collapsing
model here is isothermal and does not include rotation and magnetic field,
the estimates here should be considered as references.

% in Froebrich2003, Lb= 4.5 Lsun d=315pc, here d=321 pc
% Since L propto d^2 ==> adjusted to 4.67 Lsun with the new distance.

%calc "0.23e5**3/6.67e-8*3.156e7/2e33" == 2.88e-6

% at the outer edge and at 10$^4$ yr age the outer edge has reached 1000 au
%where the rotation veloicty is too small to have any noticable effect on
%the collapse.

% In the inside-out collapsing model, the centrifugal radius where that material
% is l^2/GM = 0.0579 cs* Omega^2 t^3, or increasing with t^3
% however, it does not include rotation and magnetic field.
% 

% R_sun 6.955e10 cm

% calc "6.67e-8*0.08*2e33/4e33*2.25e-6*2e33/3.156e7/(2*6.955e10)" = 2.73 Lsun

% Infall rate assuming subtending 10 degree  == (4*pi*10/180)
% At 100 au, vinf = 1.06 km/s
% calc "1.62e-15*(100*1.5e13)**2*3.156e7*1.06e5/2e33*(4*pi*10/180)" = 4.256e-6

% Using d=280 pc in Tanner2001

We can estimate the infall rate in the flattened envelope around the disk at
a radius $r\sim 100$ au to cross check the accretion rate.  The
mass column density there can be estimated with the continuum emission at
232 GHz.  For the flattened envelope near the disk, we assume a mass opacity
of $\kappa_\nu \sim 0.016$ cm$^2$ g$^{-1}$, which is the mean value in
protostellar cores \citep{Ossenkopf1994} and protostellar disks
\citep{Beckwith1990}.  Assuming an excitation temperature of $\sim$ 40 K
(judging from the high brightness temperature of $\sim$ 20 K there in
\cCO{}), the mass column density is estimated to be $\sim 2.43$ g \cms{}. 
Dividing this column density by a path length of $\sim$ 100 au along the
line of sight, we obtain a mass density $\rho \sim 1.62\times10^{-15}$ g
\cmc{}.  The infall velocity there is $v_i \sim$ 1.06 \vkm{} (Figure
\ref{fig:pvenv}b).  Assuming that the flattened envelope subtends an angle
of $\sim$ 10\degree{}, then the infall rate is $\sim 0.7 r^2 \rho v_i \sim
4.3\times 10^{-6}$ \solarmass{} yr$^{-1}$.  Assuming that about 70\% of this
accretes to the central protostar, then the accretion rate would be $\sim
3.0\times 10^{-6}$ \solarmass{} yr$^{-1}$.  Considering the uncertainties in
the mass opacity, temperature, and other parameters, the rough estimate here
is consistent with the above.

% isothermal spherical infall rate ~ a^3/G (Shu et al 1997)
%calc "0.23e5**3/6.67e-8*3.156e7/2e33" ~ 2.8784898E-06 Ms/yr

%calc "0.7*(100*1.5e13)**2*1.62e-15*1.06e5*3.156e7/2e33"

%1.125e24 at 30 K
%8.4e23 at 40 K
%6.754e23 at 50K
% Multiply it time 1.4*2*1.66e-24 to get the mass column density

% 4 \pi 10/180 ~ 0.698

%Beckwith 0.1 (nu/1e3)**beta    ~ 0.0232 if beta=1
%Ossenkopf 0.009 (nu/231)**beta  ~ 0.009 if beta=1

% ==> 0.016$\pm0.007

\subsection{Pseudodisk with a Toroidal and Pinched Field Morphology}

In order to understand the inferred field morphology and how it comes to be,
we compare it to the simulated field morphology resulting from a collapse of
a magnetized rotating cloud core.  In HH 211, since there is a large
misalignment of $\sim$ 60\degree{} between the magnetic field axis and the
rotation axis in the cloud core, we compare our results to those in
\citet{Hirano2019}, which assumed a misalignment of 45\degree{}.  Figure
\ref{fig:Hiranomodel} shows a schematic view of their simulation result
adopted from their paper.  In their model, the field axis ({\bf B$_0$}) in
the cloud core is initially north-south oriented, roughly the same as that
in HH 211.  The rotation axis {\bf J$_0$} in the cloud core and thus in the
resulting disk has a P.A.  of $-$45\degree{}, roughly parallel (although in
opposite) to that of the jet axis observed in HH 211, which has a P.A.  of
116.6\degree{}.  The cloud core is initially spherical and has a radius of
$\sim$ $10^4$ au, similar to that of the NH$_3$ extended envelope.  It
contracts along the magnetic-field lines towards the midplane (equatorial
plane) and forms a flattened envelope (the so-called pseudodisk).  The
flattened envelope is warped because the contraction direction of the cloud
core and thus the minor axis of the flattened envelope are changed gradually
from parallel to the magnetic field axis in the outer region to parallel to
the rotation axis in the inner region.  The inner part of the flattened
envelope is infalling towards the center and thus is threaded with an
hourglass (poloidal) field morphology with a pinch in the midplane (see
Figure \ref{fig:Hiranosim} for the inner region within $\sim$ 400 au of the
central source).  This pinched field morphology also has a toroidal
component generated by the rotation in the flattened envelope.  In this
inner part of the flattened envelope, rotation velocity increases towards
the center.  Thus, the innermost part of the flattened envelope, where the
rotation dominates the dynamics and the Ohmic dissipation weakens the
magnetic field, has transformed into a rotationally supported disk around
the angular momentum vector ({\bf J$_0$}) surrounding the central protostar. 
Outflow, jet, and knots are seen coming out from the disk, with their axes
varying with time.  The details within the disk are ignored here because we
only focus on the flattened envelope.

In HH 211, the NH$_3$ extended envelope (which is more perpendicular to the
large-scale magnetic field axis) and the \cCO{} flattened envelope (which is
more perpendicular to the rotation axis and falling towards the central
source) can be identified respectively as the outer part and the inner part
of the flattened envelope in the model.  Moreover, the inferred field
morphology in the \cCO{} flattened envelope seems roughly consistent with
the simulated field morphology in the inner part of the flattened envelope
as shown in Figure \ref{fig:Hiranosim}.  For example, a pinched field
morphology is seen threading the flattened envelope, tentative toroidal
fields are seen above and below the midplane around the outflow axis, and
poloidal fields are seen in the envelope around the outflow base.  As
mentioned earlier, a point symmetry in the polarization distribution is seen
in the flattened envelope at 232 GHz, with fewer detections on the east side
in the northeastern envelope and fewer detections on the west side in the
southwestern envelope.  Such a point symmetry has also been seen in the
simulated polarization degree map of the flattened envelope
\citep{Kataoka2012}.  Figure 10e in \citet{Kataoka2012} shows the simulated
polarization degree map for their model 2 at a similar inclination of
80\degree{}.  In that model, the pinched field morphology also has a
comparable toroidal field component generated by the rotation.  Rotating the
simulated map by 126\degree{} counterclockwise to match the observed major
axis of the flattened envelope, we find that this point symmetry matches the
observed point symmetry here in HH 211.  As discussed in
\citet{Tomisaka2011}, a magnetic field with both toroidal and hourglass
poloidal components is point symmetric.  When such a magnetic field is
viewed at an inclination angle, the low polarization region has a
point-symmetric distribution \citep{Kataoka2012}.  In addition, the fact
that almost no polarization is detected along the outflow axis slightly
above and below the central source could also be due to a depolarization
caused by the coexisting poloidal and toroidal fields, as shown in Figure
10e in \citet{Kataoka2012}.

In summary, the \cCO{} flattened envelope observed here in HH 211 can be considered
as the inner part of a pseudodisk threaded with a pinched field morphology,
with its material spiraling inwards towards the center.  It is formed likely
because the core material is guided by the field lines and falls towards the
midplane.  It is misaligned with the rotating disk likely due to the
misalignment between the axis of the magnetic field and the axis of the
rotation in the cloud core.  It is not rotationally supported and its
material is falling preferentially along the equatorial plane, producing the
pinched (poloidal) field morphology.  It is also rotating, generating the
toroidal field component in the pinched field morphology, and thus the
point-symmetric polarization distribution along the equatorial plane and
almost no polarization detection along the outflow axis slightly below and
above the source in the flattened envelope.

\subsection{Disk Formation and Magnetic Braking}

%$\sim$ $\frac{4}{3}$ times the radius of the keplerian disk
%\citep{Lee2017disk}.

The kinematics in the flattened envelope can be well described by a
ballistic infalling-rotating model.  In this model, the material is assumed
to have the same specific angular momentum.  As discussed earlier, this is
expected from the inside-out collapse, because the material in the flattened
envelope comes from a thin shell in the extended envelope and thus has
roughly the same specific angular momentum.  In addition, in this ballistic
infalling-rotating model, only a fraction ($\sim$ 30\%) of angular momentum
is lost before the flattened envelope transforms into a rotating disk around
the protostar \citep{Sakai2014,Lee2017COM}.  Thus, no significant magnetic
braking is detected in HH 211.  In this system, a large misalignment has
already been seen between the magnetic field axis and rotation axis in the
cloud core \citep{Matthews2009}, down to the inner part \citep{Hull2014},
thus the magnetic braking is indeed expected to be less efficient
\citep{Joos2012}.  This misalignment likely causes the misalignment between
the flattened envelope axis and the disk axis, promoting the disk formation
\citep{Li2014,Vaisala2019,Hirano2019}.  Since the material in the flattened
envelope needs to lose its angular momentum before joining the rotating
disk, a wind or an outflow is expected to be launched near the centrifugal
barrier.  More observations are required to check this.

%It would be better if you could comment on the centrifugal barrier and the launching
%radius of the disk wind. 

%The flattened envelope axis appears slightly misaligned with the disk axis,
%rotated counterclockwise by $\sim$ 8\degree{} from the disk axis.  This
%misalignment could be the pseudodisk warping caused by the misalignment
%between the magnetic field axis and rotation axis in the cloud core,
%promoting the disk formation \citep{Li2014,Vaisala2019,Hirano2019}.

\subsection{Comparison of disk sizes in HH 211 vs.  other Class 0 protostars}

HH 211 is one of the youngest sources found to host a rotating disk around
the protostar \citep{Lee2018HH211Disk}.  In this source, the disk
has a radius of $\sim$ 20 au, much smaller than that in HH 212 (44 au) and
L1527 (54 au), likely because the flattened envelope has a much smaller
specific angular momentum ($\sim$ 55 au \vkm{}) than that in HH 212 ($\sim$
140 au \vkm{}) \citep{Lee2017COM} and L1527 ($\sim$ 130 au \vkm{})
\citep{Ohashi2014}.  This is in turn likely because this source is much
younger, and thus only the inner part of the cloud core has collapsed
towards the center, as shown in Figure \ref{fig:rcollapse}.  Based on the
inside-out collapsing model, the flattened envelope has a specific angular
momentum $l \propto t_c^{1.55}$ (see Section \ref{sec:collapse}).  Since the
central protostellar mass grows with $M_\ast \propto t_c$, the centrifugal
radius (and thus the disk radius) will grow with $\frac{l^2}{GM_\ast}
\propto t_c^{2.1}$.  Notice that since this inside-out collapsing model is
isothermal and does not include rotation and magnetic field, the growth rate
here is only for a reference.  The actual growth rate could be slower,
as found in \citet{Lee2018HH211Disk}.

% In Basu 1998, l propto r ==> r_d \propto t
% In Terebey, l propto r^2 ==> r_d \propto t^3
% In both case, M_* propto t, as in inside-out collapse model.

\subsection{Misalignment between the Envelope and the Disk, and the Jet Axis} 

As described earlier, the major axis of the flattened envelope appears
rotated counterclockwise by $\sim$ 8\degree{} from the major axis of the
disk.  This misalignment may cause a disk warping and the disk orientation
to rotate counterclockwise when the envelope material joins the disk. 
Interestingly, the jet axis in this system has already been found to have
rotated counter-clockwise by $\sim$ 3\degree{} in the past
\citep{Eisloffel2003}, suggestive of a counter-clockwise rotation of the
disk orientation in the past.  Thus, the misalignment between the flattened
envelope and the disk may cause the jet axis continue to rotate
counter-clockwise.

%Interestingly, these orientations are also roughly the same as those seen
%in a larger scale in molecular cloud at 345 GHz (850 \micron)
%\citep{Matthews2009} and in envelope scale at 230 GHz at $\sim$
%\arcsa{4}{1} resolution \citep{Hull2014}.

%Some small rotation in polarization near the center from 350
%GHz to 230 GHz could be due to the beam effect, because the 230 GHz is rotated.

%Note that impol by default will remove the bias when calculating polarized
%intensity, i.e., P ~ sqrt(Q^2+U^2 - sigma_p^2), where sigma_p is provided
%by user, P_fraction = P / I. Note that only when Q^2+U^2 > sigma_p^2.

%Since the thermal emission (with a brightness temperature of a few K and a
%peak of $\sim$ 22 K at the center at 350 GHz, and 29 K at 230 GHz) of the
%dust is optically thin, the implied magnetic field is perpendicular to the
%polarization orientation, aligned with that seen in the larger (outer)
%envelope scale, neither parallel and perpendicular to the disk axis.

%The spectral index is , and thus could indicate the big grain in the disk. 
%HOwever, as discussed in HL Tau, altough most emission is from mm grain, the
%dust polarization is from 100 um grain. Distribution are different.

\subsection{Magnetic Field Strength}

In the flattened envelope, a toroidal field component is required in the
pinched field morphology to produce the observed point symmetric
polarization distribution.  Since the toroidal field is generated by the
rotation in the envelope, the magnetic energy should be comparable to the
rotation energy.  Thus, by equating the magnetic energy to the rotation
energy, we have the field strength to be $B = \sqrt{4\pi \rho} v_\phi$. 
Here, we estimate the field strength in the envelope around the disk at a
distance of 100 au from the center.  The rotation velocity there is $v_\phi
\sim$ 0.55 \vkm{} (Figure \ref{fig:pvenv}b) and the mass density there is
$\rho \sim 1.62\times10^{-15}$ g \cmc{} (see Section \ref{sec:collapse}). 
Thus, the field strength is estimated to be $ \sim 7.8 $ mG at a distance of
100 au from the central protostar.  In spite of this strong field around the
disk, there is no clear sign of an efficient magnetic braking, likely
because of the large misalignment between the rotation axis and magnetic
field axis in the cloud core.

Here we check if the field strength in HH 211 is reasonable by
comparing it to that estimated recently in the Class 0 source L1448-IRS2. 
In that source, \citet{Kwon2019} has estimated a mean field strength of only
$\sim$ 1 mG in the envelope, using the Davis-Chandrasekhar-Fermi
relationship.  This mean field strength was derived with a mean density of
$\sim 2.5\times10^{-17}$ g \cmc{} averaged over the whole envelope, which
has a size of $\sim$ \arcs{6} (i.e., 1930 au) in the major axis and $\sim$
\arcs{3} (i.e., 963 au) in the minor axis.  This density is a factor of
$\sim$ 65 lower than that derived here in HH 211 around the disk.  In the
innermost envelope as that seen here in HH 211, the density and the field
strength are both expected to be higher.  If the density there is as high as
that derived here in HH 211 and the magnetic field strength can scale with
the square root of the density as found in molecular clouds
\citep{Crutcher2012}, then the field strength could be $\sim$ 8mG, and thus similar
to that derived here in HH 211 around the disk.  Interestingly, the rotation
velocity there could also be similar to that measured here in HH 211,
judging from the previously reported rotation velocity in the envelope of
L1448-IRS2 \citep{Tobin2015}.

% assuming a dynamical infall with rho \propto r^{-1.5} then rho = rho_m (r/r_m)^(-1.5)
% where rho_m is the mean density over a radius of r_m
% for example in L1448 irs2, we have rho ~ 2.5e-17 (r/3")^(-1.5), assuming r_m = 3"

%Assuming a magnetic flux frozen in and the $B\propto r^{-2}$ and thus the
%field strength at 1000 au (radius of dynamic collapse) is $\sim$ 100
%$\mu$G.

%The isothermal sound speed there is $\sim 0.38$ \vkm{}.

%is $0.189\sqrt{40/10}=0.38$ \vkm{}.

% isothermal sound speed in 0.189\sqrt(40/10)

%B^2/8pi = 1/2 rho v^2 ==> B = v_\phi sqrt(4 pi rho)

%rho = 1.4 mH2 * nH2 
%nH2 = calc "1.62e-15/(1.4*2*1.66e-24)" = 3.4853699E+08

%calc "sqrt(4*pi*1.62e-15)*0.55e5" = 7.8473883E-03

\section{Conclusions}

We have mapped the central region of the HH 211 system in continuum and
\cCO{} line.  A flattened envelope is seen in the continuum with a major
axis slightly rotated counterclockwise from the disk major axis, extending
out to $\sim$ 400 au from the outer edge of the disk.  Based on our
kinematic study in \cCO{} with a ballistic infalling-rotating model, it is
spiraling (infalling and rotating) inwards towards the center and is
expected to transform into a rotating disk at $\sim$ 20 au, consistent with
the disk radius estimated before.  In this ballistic infalling-rotating
model, the central protostar has a mass of $\sim$ 0.08 \solarmass{} and the
flattened envelope has roughly the same specific angular
momentum of $\sim$ 55$\pm15$ au \vkm{}.  This same specific angular momentum
in the flattened envelope is expected if the flattened envelope came from a thin
shell of the NH$_3$ extended envelope through an inside-out collapse.  Based
on the inside-out collapse, the collapsing age is $\sim$ 35000 yrs in HH
211.

The polarization towards the disk is spatially unresolved and could be
affected by dust self-scattering.  In the flattened envelope, the
polarization is likely due to the magnetically aligned dust grains,
inferring a highly pinched poloidal field morphology threaded on both sides
of the flattened envelope.  Thus, both the kinematics and field morphology
support that the flattened envelope is a pseudodisk formed as the infalling
gas is guided by the field lines to the equatorial plane.  In addition, a
point symmetric polarization distribution is seen in the flattened envelope,
implying that the pinched field lines also have a toroidal component
generated by the rotation in the flattened envelope.  No significant loss of
angular momentum and thus no clear magnetic braking are detected in the
flattened envelope around the disk probably because of the large
misalignment between the axis of the magnetic field and the axis of the
rotation in the cloud core.  Such a misalignment also causes a misalignment
between the flattened envelope axis and the disk axis, promoting the disk
formation.

\acknowledgements

We thank the referee for constructive comments. We thank Shingo Hirano for
providing Figures \ref{fig:Hiranomodel} and \ref{fig:Hiranosim}, showing
their simulation results to be compared with our observations.  This paper
makes use of the following ALMA data: ADS/JAO.ALMA\#2016.1.00017.S.  ALMA is
a partnership of ESO (representing its member states), NSF (USA) and NINS
(Japan), together with NRC (Canada), NSC and ASIAA (Taiwan), and KASI
(Republic of Korea), in cooperation with the Republic of Chile.  The Joint
ALMA Observatory is operated by ESO, AUI/NRAO and NAOJ.  C.-F.L. 
acknowledges grants from the Ministry of Science and Technology of Taiwan
(MoST 107-2119-M-001-040-MY3) and the Academia Sinica (Career Development
Award and Investigator Award).  Woojin Kwon was supported by Basic Science
Research Program through the National Research Foundation of Korea
(NRF-2016R1C1B2013642).

%% Remember to include "(" and ")" for the year,e.g., (1998)
%%
\def\nat{Natur}

\begin{figure} [!hbp]
\centering
\putfig{0.73}{270}{f3.eps} %{cont_c18o.eps}
\figcaption[]
{\cCO{} J=2-1 maps on top of the gray image of the continuum map at 232 GHz,
with (a) showing the integrated intensity map integrated from $\Voff=-2.7$ to 2.7 \vkm{}, (b) 
showing the blueshifted emission map integrated
from $\Voff=-2.7$ to 0 \vkm{}, and (c) showing the redshifted intensity map
integrated from $\Voff=0$ to 2.7 \vkm{}.
The continuum map at 232 GHz is the same as in Figure \ref{fig:contbeta}b.
The contours in the \cCO{} maps start at 3$\sigma$ with a step of 2$\sigma$, where $\sigma=3$ \mJybk{} 
in (a) and 2 \mJybk{} in (b) and (c).
The asterisk,
the cross, the magenta lines, and the red and blue arrows
have the same meanings as in Figure \ref{fig:contbeta}.
\label{fig:ccoenv}}
\end{figure}

\begin{figure} [!hbp]
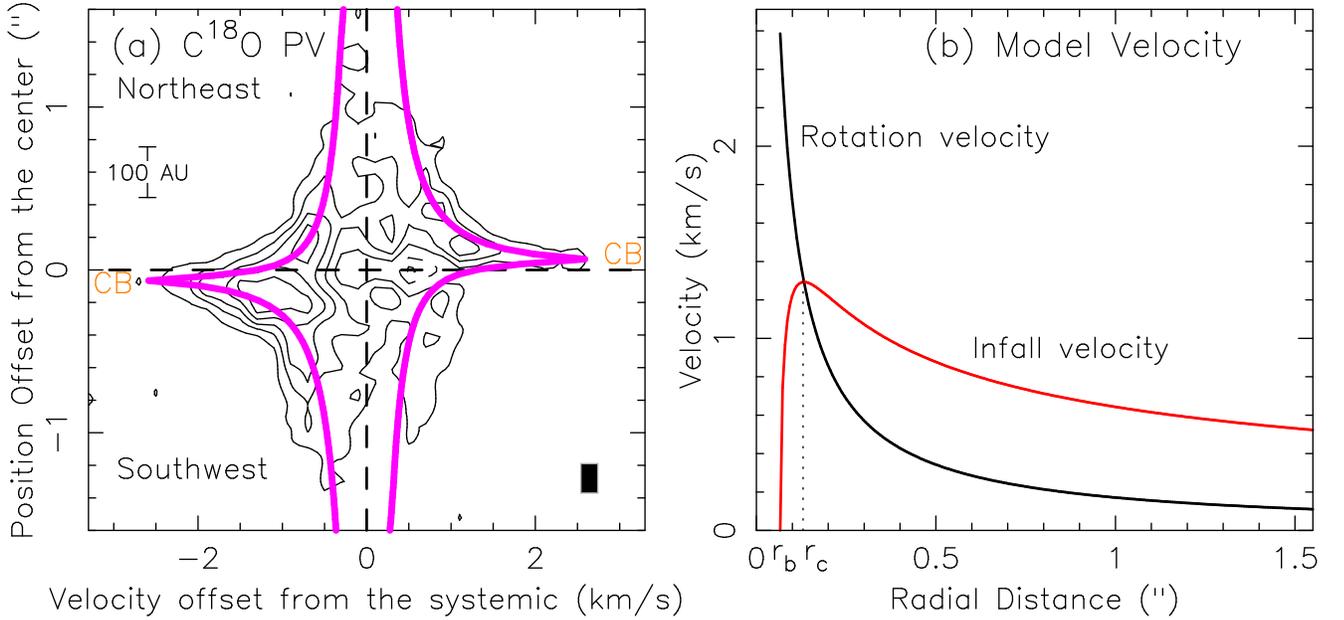

\centering
\putfig{0.7}{270}{f4.eps} %{pvc18o.eps}
\figcaption[]
{Position-Velocity (PV) diagram of the \cCO{} J=2-1 emission along the major
axis of the envelope and the model velocities to fit the PV structure (see
text).  CB stands for centrifugal barrier. 
The black rectangle in the lower-right-hand corner shows the spatial and velocity
resolutions.
The contours start at 3$\sigma$ with a step of 3$\sigma$, where
$\sigma=1.2$ \mJyb{} (i.e., 1.626 K). 
The intensity peak is $\sim$ 21 \mJyb{} (28.5 K).
The magenta curves
are the outer boundaries of the PV structure derived from the ballistic infalling-rotating
model. In panel (b), $r_c$ is centrifugal radius and $r_b$ is centrifugal barrier.
\label{fig:pvenv}}
\end{figure}

\begin{figure} [!hbp]
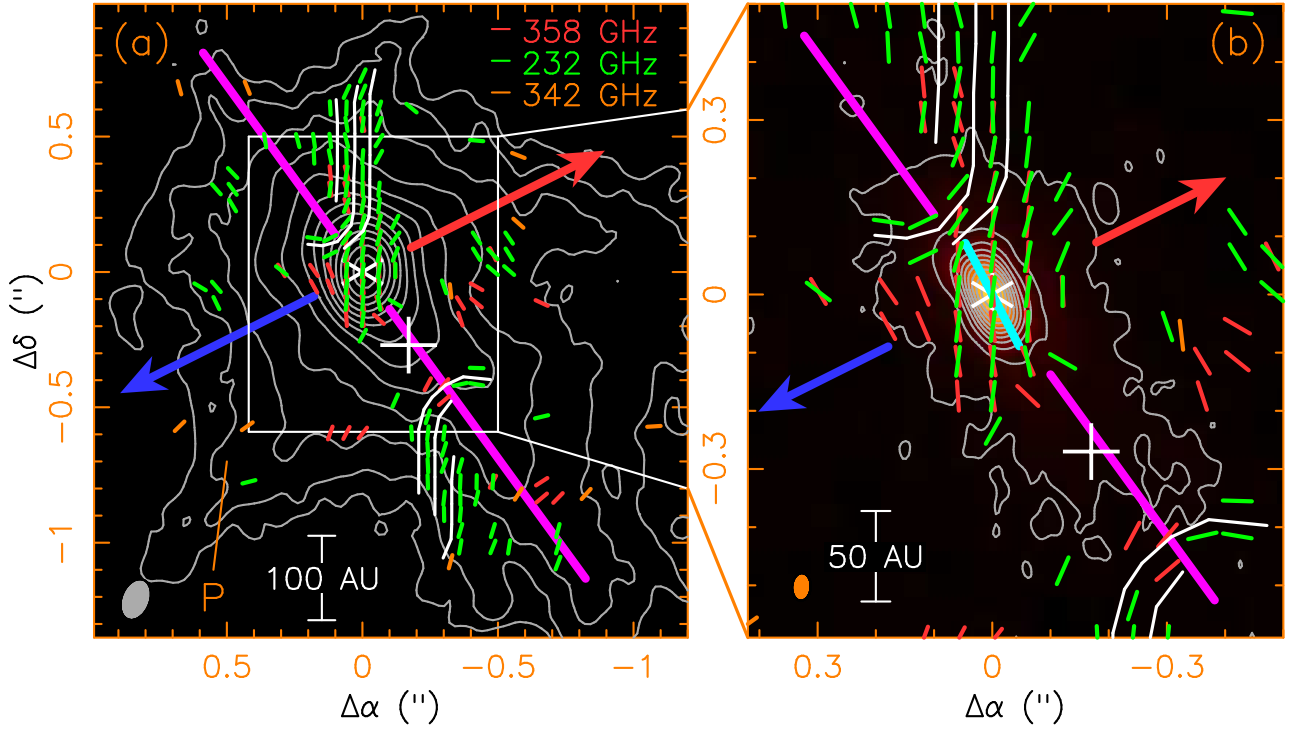

\centering
\putfig{1.1}{270}{f5.eps} %{cont_pol_B.eps}
\figcaption[]
{Magnetic field morphology of the flattened envelope in HH 211 inferred
from dust polarization. It is delineated by the line segments obtained from
rotating the polarization orientations by 90\degree{}, with red observed at
358 GHz (ALMA Band 7), green at 232 GHz (ALMA Band 6), and orange at 342 GHz
\cite[SMA results from][]{Lee2014Mag}. The asterisk, the
cross, the magenta lines, the red and blue arrows, and the cyan lines have
the same meanings as in Figure \ref{fig:contbeta}. The white curves
guide the readers for the pinched field morphology in the flattened envelope. 
``P" means poloidal field. The contours are the continuum emission map 
at 232 GHz, the same as
in Figure \ref{fig:contpol}b.  Panel (b) shows the magnetic field morphology
in the inner part, plotted on top of the disk and innermost envelope adopted
from Figure 1a of \citet{Lee2018HH211Disk} at a resolution of
\arcsa{0}{047}$\times$\arcsa{0}{031}.  The contours start at 5$\sigma$ with
a linear step of 12$\sigma$, where $\sigma=0.73$ K.
Notice that in the disk, since the
polarization can be significantly affected by dust self-scattering, the line segments there
unlikely trace the field morphology (see text).
\label{fig:Bfield}}
\end{figure}

\begin{figure} [!hbp]
\centering
\putfig{0.7}{270}{f6.eps} %{nh3.eps}
\figcaption[]
{The NH$_3$(1,1) intensity map of the extended envelope in HH 211, 
adopted from \citet{Wiseman2001}.
The contours start from 10 \mJyb{} \vkm{} with a step of 10 \mJyb{} \vkm{}.
The line segments are the magnetic field orientations from \citet{Matthews2009} shown in \citet{Hull2014}.
The asterisk marks the source position. The dot-dashed circle marks the possible boundary of the
collapsing region as discussed in the text.
The red and blue arrows have the same meanings as in Figure \ref{fig:contbeta}.
\label{fig:nh3}}
\end{figure}

\begin{figure} [!hbp] \centering \putfig{0.63}{270}{f7.eps}
 \figcaption[] {Distribution of specific angular
momentum in the \cCO{} flattened envelope (horizontal bar with
its thickness representing the uncertainty) and the
NH$_3$ extended envelope \cite[data points with error bars adopted from][updated for the new
distance of 321 pc]{Tanner2011}.  
The \cCO{} flattened envelope
is found have a roughly constant specific
angular momentum of $\sim 55\pm15$ au \vkm{} up to a radius of $\sim$ 400
au, as discussed in Section \ref{sec:env}.
The solid curve shows a fit to the
distribution of specific angular momentum in the NH$_3$ extended envelope. 
The 3 dotted curves show the distribution of specific angular momentum in
the collapsing envelope from an inside-out collapsing model (with the NH$_3$
extended envelope as an input) at 3 different ages.   The small thick line
near $\sim$ 830 au shows the location of the thin shell in the extended
envelope that forms the observed \cCO{}
flattened envelope at $t_c \sim 35000$ yrs.  
\label{fig:rcollapse}}
\end{figure}

\begin{figure} [!hbp]
\centering
\putfig{0.4}{0}{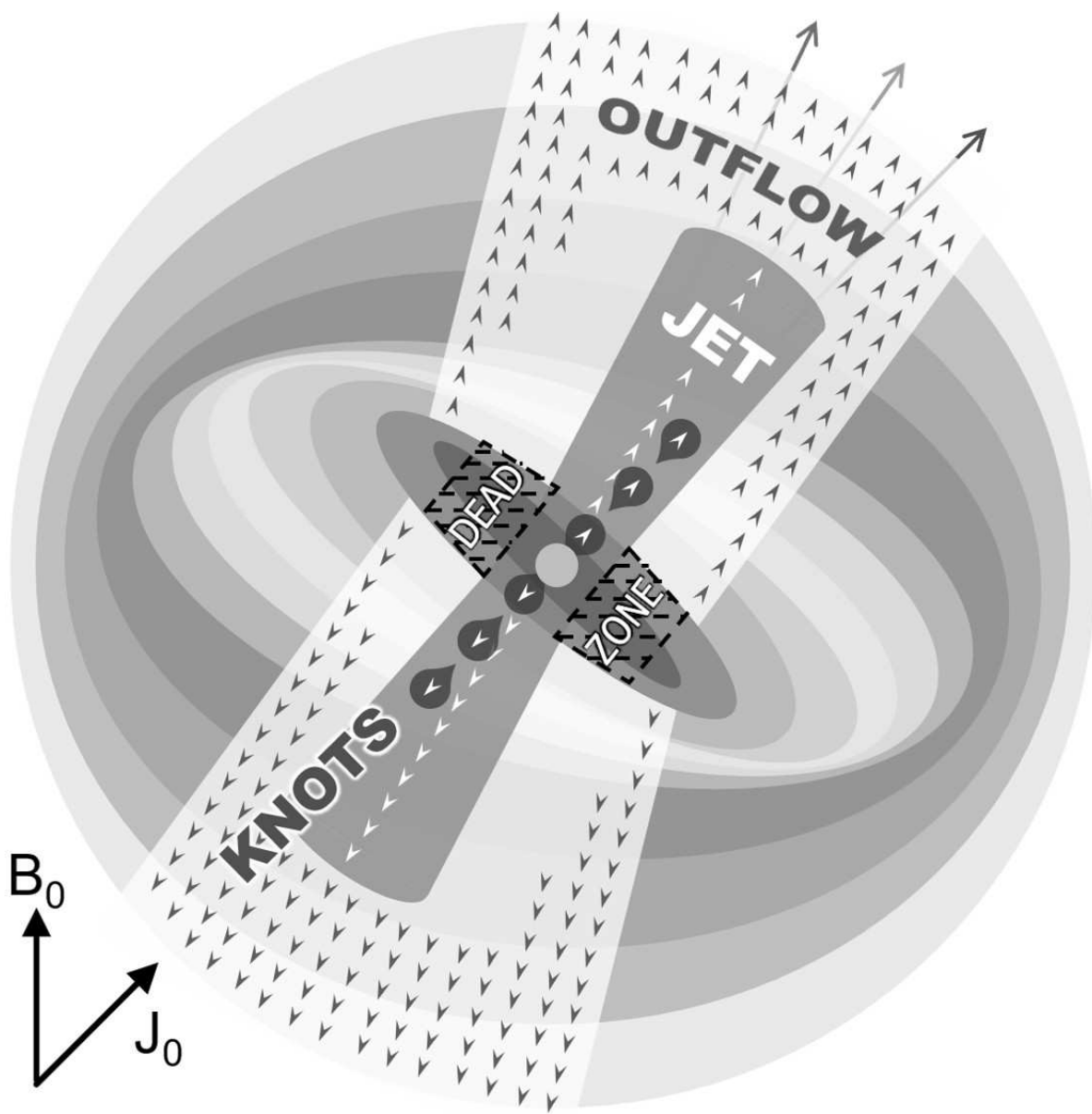} %Hirano2019_figure6.eps}
\figcaption[]
{Schematic view of the simulation result from a collapse of a magnetized
rotating cloud core in \citet{Hirano2019}.  The initial directions of the
B-field ({\bf B$_0$}) and angular momentum vector ({\bf J$_0$}) are inclined
at 45\degree{}.  The axis of the flattened envelope gradually changes
during the gravitational contraction from the outer region to the inner region.
The innermost region of the flattened envelope has transformed
into a rotating disk. Outflow, jet, and
knots are seen coming out from the disk, with their axes varying with time. 
\label{fig:Hiranomodel}}
\end{figure}

\begin{figure} [!hbp]
\centering
\putfig{0.3}{0}{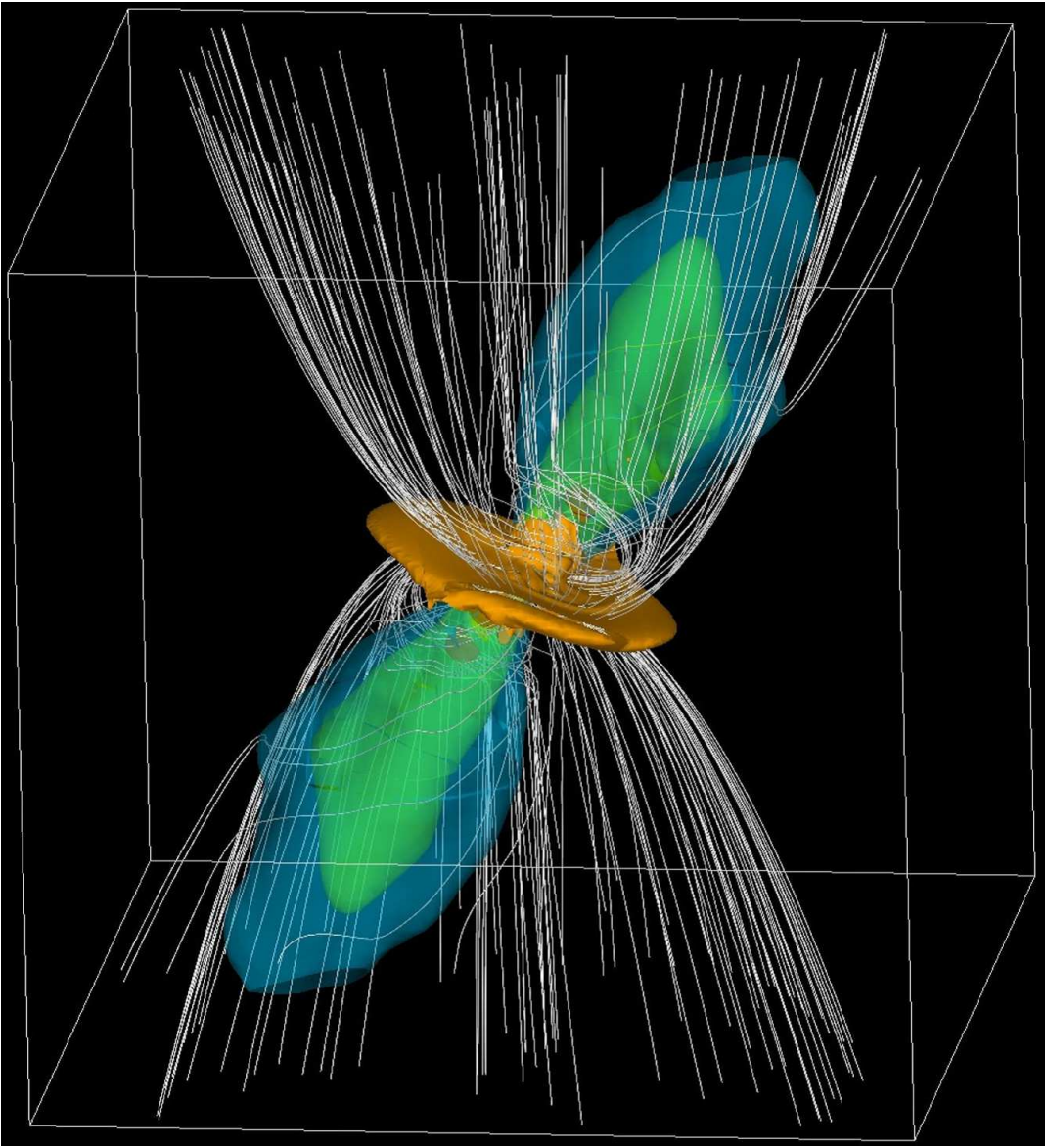}
\figcaption[]
{Model magnetic field morphology in the inner region, kindly provided
by Shingo Hirano using their results in \citet{Hirano2019} to be compared
with our observation qualitatively. The box has a size of $\sim$ 800 au. 
The orange iso-density surfaces indicate the flattened envelope.  The white lines show
the magnetic field lines, which is roughly aligned with the initial
direction at large scale but twisted at small scale because of the coupling
to the accreting gas.  The blue, green, and light orange iso-velocity
surfaces depict three outflows (low-velocity outflow, jet, and knots) 
coming out from the disk at the center.
\label{fig:Hiranosim}}
\end{figure}

%on the $y=0$ cutting plane.

\begin{table}
\small
\centering
\caption{Correlator Setup For Band 7}
\label{tab:corrb7}
\begin{tabular}{llllll}
\hline
Spectral  & Line or   & Number of & Central Frequency & Bandwidth & Channel Width\\
Window & Continuum & Channels  & (GHz)             & (MHz)     & (kHz) \\
\hline\hline
0 & SO \SOt      & 480   & 346.533 & 117.187  & 244.141  \\
1 & CO J=3-2      & 480  & 345.800 & 117.187  & 244.141  \\
2 & SiO J=8-7     & 960  & 347.335 & 468.750  & 488.281  \\
3 & HCO$^+$ J=4-3 & 960  & 356.739 & 234.375  & 244.141  \\
4 & Continuum     & 960  & 358.005 &1875.000  & 1953.125  \\
\hline
%\multicolumn{6}{l}{a: Effective Velocity Resolution is about 2 Channels in this Cycle}
\end{tabular}
\end{table}

\begin{table}
\small
\centering
\caption{Correlator Setup For Band 6}
\label{tab:corrb6}
\begin{tabular}{llllll}
\hline
Spectral  & Line or   & Number of & Central Frequency & Bandwidth & Channel Width\\
Window & Continuum & Channels  & (GHz)             & (MHz)     & (kHz) \\
\hline\hline
0 & SiO $J=5-4$         & 960  & 217.102 & 468.750  & 488.281  \\
1 & $^{13}$CO $J=2-1$   & 480  & 220.407 &  58.593  & 122.070  \\
2 & C$^{18}$O $J=2-1$   & 480  & 219.569 &  58.593  & 122.070  \\
3 & N$_2$D$^+$ $J=3-2$  & 480  & 231.328 &  58.593  & 122.070  \\
4 & CO $J=2-1$          & 480  & 230.547 &  58.593  & 122.070  \\
5 & Continuum           & 960  & 232.521 &1875.000  & 1953.125  \\
\hline
%\multicolumn{6}{l}{a: Effective Velocity Resolution is about 2 Channels in this Cycle}
\end{tabular}
\end{table}


\begin{thebibliography}{}
%\bibitem[Adande et al.(2013)]{Adande2013} Adande, G.~R., Woolf, N.~J., \&
%Ziurys, L.~M.\ 2013, AsBio, 13, 439

%\bibitem[Agra-Amboage et al.(2011)]{Agra-Amboage2011} Agra-Amboage, V.,
%Dougados, C., Cabrit, S., \& Reunanen, J.\ 2011, \aap, 532, A59

\bibitem[Agurto-Gangas et al.(2019)]{Agurto2019} Agurto-Gangas, C., Pineda,
J.~E., Sz{\"u}cs, L., et al.\ 2019, \aap, 623, A147


\bibitem[Allen, Li, \& Shu(2003)]{Allen2003} Allen, A., Li, Z.,
 \& Shu, F.~H.\ 2003, \apj, 599, 363
%\bibitem[Alten et al.(1997)]{Alten1997} 
% Alten, V. P., Bally, J., Devine, D. and Miller, G. J. 1997, IAU Symp. 182:
% Herbig-Haro Flows and the Birth of Stars, 182, 51P
%\bibitem[Anderson et al.(2003)]{Anderson2003} Anderson, J.~M., Li, 
%Z.-Y., Krasnopolsky, R., \& Blandford, R.~D.\ 2003, \apjl, 590, L107 


%\bibitem[Andersson et al.(2015)]{Andersson2015} Andersson, B.-G., Lazarian,
%A., \& Vaillancourt, J.~E.\ 2015, \araa, 53, 501

%\bibitem[Andr\'e \& Montmerle(1994)]{Andre1994} Andr\'e, P.  \& 
%  Montmerle, T. 1994, \apj, 420, 837 
%\bibitem[Andre et al.(2000)]{Andre2000} Andre, P., Ward-Thompson, 
%D., \& Barsony, M.\ 2000, Protostars and Planets IV, 59  
%\bibitem[Andrews 
%\& Williams(2007)]{Andrews2007} Andrews, S.~M., \& Williams, J.~P.\
%2007, \apj, 659, 705 

%\bibitem[Andrews et al.(2009)]{Andrews2009} Andrews, S.~M., Wilner, 
%D.~J., Hughes, A.~M., Qi, C., \& Dullemond, C.~P.\ 2009, \apj, 700, 1502 

%\bibitem[Anglada \& Rodr{\'{\i}}guez(2002)]{Anglada2002} Anglada, 
%G.~\& Rodr{\'{\i}}guez, L.~F.\ 2002, Revista Mexicana de Astronomia y 
% Astrofisica, 38, 13 





%\bibitem[Arce \& Sargent(2004)]{Arce2004} Arce, H.~G.~\& 
% Sargent, A.~I.\ 2004, \apj, 612, 342 
%\bibitem[Arce et al.(2007)]{Arce2007} Arce, H.~G., Shepherd, D., 
%Gueth, F., Lee, C.-F., Bachiller, R., Rosen, A., 
%\& Beuther, H.\ 2007, Protostars and Planets V, 245 

\bibitem[Bacciotti et al.(2018)]{Bacciotti2018} Bacciotti, F., Girart, J.~M., Padovani, M., et al.\ 2018, \apjl, 865, L12 


%\bibitem[Bachiller(1996)]{Bachiller1996}Bachiller, R. 1996, ARAA, 34, 111
%\bibitem[Bachiller et al.(1995)]{Bachiller1995} Bachiller, R., 
% Guilloteau, S., Dutrey, A., Planesas, P. \& Martin-Pintado, J. 1995, \aap, 
% 299, 857 

%\bibitem[Balbus \& Hawley(1991)]{Balbus1991} Balbus, S.~A., \& Hawley, J.~F.\ 1991, \apj, 376, 214 


%\bibitem[Balbus \& Hawley(2006)]{Balbus2006} Balbus, S.~A., \& Hawley,
%J.~F.\ 2006, \apj, 652, 1020


%\bibitem[Bally \& Lada(1983)]{Bally1983} Bally, J. \& Lada, C. J. 
% 1983, \apj, 265, 824 
\bibitem[Beckwith et al.(1990)]{Beckwith1990} Beckwith, S.~V.~W., 
Sargent, A.~I., Chini, R.~S., \& Guesten, R.\ 1990, \aj, 99, 924 
%\bibitem[Belloche \& Andr{\' e}(2004)]{Belloche2004} Belloche, A.~\& 
% Andr{\' e}, P.\ 2004, \aap, 419, L35 
%\bibitem[Bence, Richer \& Padman(1996)]{Bence1996} Bence, S. J., 
%  Richer, J. S. \& Padman, R. 1996, \mnras, 279, 866 
%\bibitem[Bence et al.(1998)]{Bence1998} Bence, S. J., Padman, R., 
% Isaak, K. G., Wiedner, M. C. \& Wright, G. S. 1998, \mnras, 299, 965 


%\bibitem[Bjerkeli et al.(2016)]{Bjerkeli2016} Bjerkeli, P., van der Wiel,
%M.~H.~D., Harsono, D., Ramsey, J.~P., \& J{\o}rgensen, J.~K.\ 2016, \nat,
%540, 406




%\bibitem[Bohigas, Persi \& Tapia(1993)]{Bohigas1993} Bohigas, J., 
% Persi, P. \& Tapia, M. 1993, \aap, 267, 168 

%\bibitem[Brauer et al.(2008)]{Brauer2008} Brauer, F., Dullemond, C.~P., \& Henning, T.\ 2008, \aap, 480, 859 



%\bibitem[Cabrit, Raga \& Gueth(1997)]{Cabrit1997} Cabrit, S., Raga, A., Gueth, F.
%  1997, IAUS, 182, 163
%\bibitem[Cabrit et al.(2007)]{Cabrit2007} Cabrit, S., Codella, C., 
%Gueth, F., Nisini, B., Gusdorf, A., Dougados, C., \& Bacciotti, F.\ 2007, \aap, 468, L29 
%\bibitem[Cabrit et 
%al.(2012)]{Cabrit2012} Cabrit, S., Codella, C., Gueth, F., \& Gusdorf, A.\ 2012, \aap, 548, L2 
%\bibitem[Caselli, Myers, \& Thaddeus(1995)]{Caselli1995} Caselli, 
% P., Myers, P.~C., \& Thaddeus, P.\ 1995, \apjl, 455, L77 
%\bibitem[Caselli et al.(1997)]{Caselli1997} Caselli, P., Hartquist, 
%T.~W., \& Havnes, O.\ 1997, \aap, 322, 296 
%\bibitem[Caselli et al.(2002)]{Caselli2002} Caselli, P., Benson, 
%P.~J., Myers, P.~C., \& Tafalla, M.\ 2002, \apj, 572, 238

%\bibitem[Ceccarelli et al.(2001)]{Ceccarelli2001} Ceccarelli, C., Loinard,
%L., Castets, A., et al.\ 2001, \aap, 372, 998

%\bibitem[Cerqueira et al.(2006)]{Cerqueira2006} Cerqueira, A.~H.,
%Vel{\'a}zquez, P.~F., Raga, A.~C., Vasconcelos, M.~J., \& de Colle, F.\
%2006, \aap, 448, 231

%\bibitem[Chapman et al.(2002)]{Chapman2002} Chapman, N.~L., Mundy, L.~G.,
%Lee, C.-F., \& White, S.~M.\ 2002, Bulletin of the American Astronomical
%Society, 34, 1133

\bibitem[Chapman et al.(2013)]{Chapman2013} Chapman, N.~L., Davidson, J.~A.,
Goldsmith, P.~F., et al.\ 2013, \apj, 770, 151

%\bibitem[Chernin et al.(1994)]{Chernin1994} Chernin, L.M., Masson, C.R.,
%  Gouveia dal Pino, E.M., Benz, W., 1994, ApJ, 426, 204
%\bibitem[Chini et al.(1997)]{Chini1997} Chini, R., Reipurth, B., 
%Sievers, A., Ward-Thompson, D., Haslam, C.~G.~T., Kreysa, E., \& Lemke, R.\ 
%1997, \aap, 325, 542 
%\bibitem[Cernicharo \& Reipurth(1996)]{Cernicharo1996} Cernicharo, J. 
%  \& Reipurth, B. 1996, \apjl, 460, L57 


%\bibitem[Charnley(1997)]{Charnley1997} Charnley S.B. 1997, ApJ 481, 396

%\bibitem[Choi et al.(2010)]{Choi2010} Choi, M., Tatematsu, K., 
%\& Kang, M.\ 2010, \apjl, 723, L34 
%\bibitem[Chrysostomou et al.(2008)]{Chrysostomou2008} Chrysostomou, A., 
%Bacciotti, F., Nisni, B., Ray, T.~P., Eisloffel, J., Davis, C.~J., 
%\& Takami, M. \ 2008, \aap, 482, 575 
%\bibitem[Claussen et al.(1998)]{Claussen1998}Claussen, M.J., Marvel., K.B., 
%  Wootten, A., Wilking, B.A. 1998, ApJL, 507, L79
%\bibitem[Codella et al.(2005)]{Codella2005} Codella, C., Bachiller, 
%R., Benedettini, M., Caselli, P., Viti, S., \& Wakelam, V.\ 2005, \mnras, 
%361, 244 


%\bibitem[Codella et al.(2007)]{Codella2007} Codella, C., Cabrit, S., Gueth, F., et al.\ 2007, \aap, 462, L53 

%\bibitem[Codella et al.(2014)]{Codella2014} Codella, C., Cabrit, S., Gueth, F., et al.\ 2014, \aap, 568, L5 

%\bibitem[Coffey et al.(2007)]{Coffey2007} Coffey, D., Bacciotti, 
%F., Ray, T.~P., Eisl{\"o}ffel, J., \& Woitas, J.\ 2007, \apj, 663, 350 
%\bibitem[Cohen(1980)]{Cohen1980} Cohen, M. 1980, \aj, 85, 29 
%\bibitem[Coppin, Davis \& Micono(1998)]{Coppin1998} Coppin, K. E. 
%  K., Davis, C. J. \& Micono, M.  1998, \mnras, 301, L10 

\bibitem[Cox et al.(2018)]{Cox2018} Cox, E.~G., Harris, R.~J., Looney, L.~W., et al.\ 2018, \apj, 855, 92 

\bibitem[Crutcher(2012)]{Crutcher2012} Crutcher, R.~M.\ 2012, \araa, 50, 29 




%\bibitem[Davis, Mundt \& Eisloeffel(1994)]{Davis1994} Davis, C. J., 
%  Mundt, R.  \& Eisloeffel, J.  1994, \apjl, 437, L55 
%\bibitem[Davis et al.(1997)]{Davis1997} Davis, 
%  C. J., Ray, T. P., Eisloeffel, J. \& Corcoran, D. 1997, \aap, 324, 263 
%\bibitem[Davis et al.(2000)]{Davis2000} Davis, C.~J., Berndsen, 
%A., Smith, M.~D., Chrysostomou, A., \& Hobson, J.\ 2000, \mnras, 314, 241 
%\bibitem[de Geus, Bronfman, \& Thaddeus(1990)]{deGeus1990} de 
% Geus, E.~J., Bronfman, L., \& Thaddeus, P.\ 1990, A\&A, 231, 137 
%\bibitem[Delamarter, Frank, \& Hartmann(2000)]{Delamarter2000} 
% Delamarter, G., Frank, A., \& Hartmann, L.\ 2000, \apj, 530, 923
%\bibitem[Dent et al.(2003)]{Dent2003} Dent, W.~R.~F., Furuya, 
%R.~S., \& Davis, C.~J.\ 2003, \mnras, 339, 633 
%\bibitem[Downes \& Ray(1999)]{Downes1999} Downes, T. P.  Ray, T. P. 
%1999, \aap, 345, 977 

%\bibitem[Dullemond \& Dominik(2004)]{Dullemond2004} Dullemond, C.~P., \&
%Dominik, C.\ 2004, \aap, 417, 159



%\bibitem[Dutrey et al.(2011)]{Dutrey2011} Dutrey, A., Wakelam, V., Boehler,
%Y., et al.\ 2011, \aap, 535, A104


\bibitem[Eisl{\"o}ffel et al.(2003)]{Eisloffel2003} Eisl{\"o}ffel, J., Froebrich, D., Stanke, T., \& McCaughrean, M.~J.\ 2003, \apj, 595, 259 




%\bibitem[Evans(1999)]{Evans1999} Evans, N.~J., II 1999, \araa, 37, 311 
%\bibitem[Fendt \& Zinnecker(1998)]{Fendt1998} Fendt, C., \& 
%Zinnecker, H.\ 1998, \aap, 334, 750 

%\bibitem[Fern{\'a}ndez-L{\'o}pez et al.(2016)]{Fernandez2016} Fern{\'a}ndez-L{\'o}pez, M., Stephens, I.~W., Girart, J.~M., et al.\ 2016, \apj, 832, 200 




%\bibitem[Frerking et al.(1982)]{Frerking1982} Frerking, M.~A., 
% Langer, W.~D., \& Wilson, R.~W.\ 1982, \apj, 262, 590 

\bibitem[Froebrich et al.(2003)]{Froebrich2003} Froebrich, D., Smith, M.~D.,
Hodapp, K.-W., \& Eisl{\"o}ffel, J.\ 2003, \mnras, 346, 163




%\bibitem[Fukui et al.(1993)]{Fukui1993} Fukui, Y. , Iwata, T. , 
% Mizuno, A. , Bally, J.  \& Lane, A. P. 1993, Protostars and planets III 
% (A93-42937 17-90), p. 603-639., 603 


%\bibitem[Fuente et al.(2010)]{Fuente2010} Fuente, A., Cernicharo, J., Ag{\'u}ndez, M., et al.\ 2010, \aap, 524, A19 


%\bibitem[Furuya \& Aikawa(2014)]{Furuya2014} Furuya, K., \& Aikawa,
%Y.\ 2014, \apj, 790, 97

%\bibitem[Galli \& Shu(1993)]{Galli1993} Galli, D., \& Shu, F.~H.\ 
% 1993, \apj, 417, 243

 
%\bibitem[Galv{\' a}n-Madrid et al.(2004)]{Galvan2004} Galv{\' 
%a}n-Madrid, R., Avila, R., \& Rodr{\'{\i}}guez, L.~F.\ 2004, Revista 
%Mexicana de Astronomia y Astrofisica, 40, 31 
%\bibitem[Garnavich et al.(1997)]{Garnavich1997} 
%  Garnavich, P. M., Noriega-Crespo, A. , Raga, A. C. \& Bohm, K. -H.  1997, 
%  \apj, 490, 752


%\bibitem[Garrod \& Widicus Weaver(2013)]{Garrod2013} Garrod, R.~T., \&
%Widicus Weaver, S.~L.\ 2013, ChRv, 113, 8939



%\bibitem[Gibb et al.(2004)]{Gibb2004} Gibb, A.~G., Richer, 
%J.~S., Chandler, C.~J., \& Davis, C.~J.\ 2004, \apj, 603, 198 
%\bibitem[Girart et al.(2000)]{Girart2000} Girart, J.~M., 
% Estalella, R., Ho, P.~T.~P., \& Rudolph, A.~L.\ 2000, \apj, 539, 763 

%\bibitem[Girart et al.(2006)]{Girart2006} Girart, J.~M., Rao, R., \&
%Marrone, D.~P.\ 2006, Science, 313, 812

%\bibitem[Girart et al.(2018)]{Girart2018} Girart, J.~M., Fern{\'a}ndez-L{\'o}pez, M., Li, Z.-Y., et al.\ 2018, \apjl, 856, L27 

%\bibitem[Glassgold et al.(1991)]{Glassgold1991} Glassgold, A.~E., 
%Mamon, G.~A., \& Huggins, P.~J.\ 1991, \apj, 373, 254 
%\bibitem[de Gouveia dal Pino \& Benz (1993)]{Gouveia1993} de Gouveia 
% dal Pino, E. M. \& Benz, W.  1993, \apj, 410, 686  


\bibitem[Goodman et al.(1993)]{Goodman1993} Goodman, A.~A., Benson, P.~J.,
Fuller, G.~A., \& Myers, P.~C.\ 1993, \apj, 406, 528

%\bibitem[Gredel \& Reipurth(1994)]{Gredel1994} Gredel, R.  \&
% Reipurth, B. 1994, \aap, 289, L19 


%\bibitem[Greenhill et al.(2013)]{Greenhill2013} Greenhill, L.~J.,
%Goddi, C., Chandler, C.~J., Matthews, L.~D., \& Humphreys, E.~M.~L.\ 2013,
%\apjl, 770, L32


%\bibitem[Gottlieb, Myers, \& Thaddeus(2003)]{Gottlieb2003} Gottlieb, 
% C.~A., Myers, P.~C., \& Thaddeus, P.\ 2003, \apj, 588, 655 
%\bibitem[Gueth et al.(1997)]{Gueth1997} Gueth, F., Guilloteau, S., Dutrey, 
% A., \& Bachiller, R.\ 1997, \aap, 323, 943 
%\bibitem[Gueth et al.(1998)]{Gueth1998} Gueth, F., Guilloteau, 
%S., \& Bachiller, R.\ 1998, \aap, 333, 287 
\bibitem[Gueth \& Guilloteau(1999)]{Gueth1999} Gueth, F. \& 
  Guilloteau, S. 1999, \aap, 343, 571 
%\bibitem[Gueth, Guilloteau \& Bachiller(1996)]{Gueth1996} Gueth, 
%  F., Guilloteau, S. \& Bachiller, R. 1996, \aap, 307, 891

%\bibitem[Guillet et al.(2011)]{Guillet2011} 
%Guillet V., Pineau des For$\hat {\rm e}$ts G., \& Jones A.P. 2011, A\&A 527, 123


%\bibitem[Harsono et al.(2014)]{Harsono2014} Harsono, D., J{\o}rgensen, J.~K., van Dishoeck, E.~F., et al.\ 2014, \aap, 562, A77 
%\bibitem[Hartigan et al.(2000)]{Hartigan2000} Hartigan, P., Bally,
%J., Reipurth, B., \& Morse, J.~A.\ 2000, Protostars and Planets IV, 841
%\bibitem[Hartmann et al.(1994)]{Hartmann1994} Hartmann, L., Boss, 
% A., Calvet, N., \& Whitney, B.\ 1994, \apjl, 430, L49 
%\bibitem[Hartmann et al.(1996)]{Hartmann1996} Hartmann, L., Calvet, 
%N., \& Boss, A.\ 1996, \apj, 464, 387

%\bibitem[Hatchell et al.(1998)]{Hatchell1998} Hatchell J., Thompson M.A., Millar T.J., \&  MacDonald G.H. 1998, A\&A 338, 713


%\bibitem[Hayashi et al.(1993)]{Hayashi1993} Hayashi, M., Ohashi, 
%N., \& Miyama, S.~M.\ 1993, \apjl, 418, L71 

\bibitem[Hennebelle \& Ciardi(2009)]{Hennebelle2009} Hennebelle, P., \& Ciardi, A.\ 2009, \aap, 506, L29 


%\bibitem[Herbst \& van Dishoeck(2009)]{Herbst2009} Herbst, E., \& van
%Dishoeck, E.~F.\ 2009, \araa, 47, 427

%\bibitem[Hirano et al.(2006)]{Hirano2006} Hirano, N., Liu, S.-Y., 
%Shang, H., Ho, P.~T.~P., Huang, H.-C., Kuan, Y.-J., McCaughrean, M.~J., \& 
%Zhang, Q.\ 2006, \apjl, 636, L141 

\bibitem[Hirano \& Machida(2019)]{Hirano2019} Hirano, S., \& Machida, M.~N.\ 2019, \mnras,  




%\bibitem[Hirota et al.(2017)]{Hirota2017} Hirota, T., Machida, M.~N.,
%Matsushita, Y., et al.\ 2017, Nature Astronomy, 1, 0146



%\bibitem[Ho et al.(2004)]{Ho2004} Ho, P. T. P., Moran, J. M., \& Lo, K. Y. 
%B2004, ApJ, 616, L1
%\bibitem[Hodapp(1994)]{Hodapp1994} Hodapp, K. -W.  1994, \apjs, 94, 
% 615 
%\bibitem[Hodapp \& Ladd(1995)]{Hodapp1995} Hodapp, K. -W.  \& Ladd, 
%  E. F. 1995, \apj, 453, 715  
%\bibitem[Hogerheijde et al.(1998)]{Hogerheijde1998} Hogerheijde, M.~R., 
%van Dishoeck, E.~F., Blake, G.~A., \& van Langevelde, H.~J.\ 1998, \apj, 
%B502, 315
%\bibitem[Hogerheijde(2001)]{Hogerheijde2001} Hogerheijde, M.~R.\ 2001, 
%\apj, 553, 618

\bibitem[Hull et al.(2013)]{Hull2013} Hull, C.~L.~H., Plambeck, R.~L., Bolatto, A.~D., et al.\ 2013, \apj, 768, 159 

\bibitem[Hull et al.(2014)]{Hull2014} Hull, C.~L.~H., Plambeck, R.~L., Kwon, W., et al.\ 2014, \apjs, 213, 13 






\bibitem[Hull et al.(2018)]{Hull2018} Hull, C.~L.~H., Yang, H., Li, Z.-Y., et al.\ 2018, \apj, 860, 82 




\bibitem[Jhan \& Lee(2016)]{Jhan2016} Jhan, K.-S., \& Lee, C.-F.\ 2016, \apj, 816, 32 

\bibitem[Joos et al.(2012)]{Joos2012} Joos, M., Hennebelle, P., \& Ciardi,
A.  2012, \aap 543, A128
%\bibitem[J{\o}rgensen et al.(2004)]{Jorgensen2004} J{\o}rgensen, 
% J.~K., Sch{\" o}ier, F.~L., \& van Dishoeck, E.~F.\ 2004, \aap, 416, 603 
%\bibitem[J{\o}rgensen et al.(2007)]{Jorgensen2007} J{\o}rgensen, 
%J.~K., et al.\ 2007, \apj, 659, 479 

%\bibitem[J{\o}rgensen et al.(2005)]{Jorgensen2005} J{\o}rgensen, J.~K.,
%Sch{\"o}ier, F.~L., \& van Dishoeck, E.~F.\ 2005, \aap, 437, 501

%\bibitem[J{\o}rgensen et al.(2016)]{Jorgensen2016} J{\o}rgensen, J.~K., van
%der Wiel, M.~H.~D., Coutens, A., et al.\ 2016, \aap, 595, A117




%\bibitem[Kahane et al.(2013)]{Kahane2013} Kahane, C., Ceccarelli, C., Faure, A., \& Caux, E.\ 2013, \apjl, 763, L38 


\bibitem[Kataoka et al.(2017)]{Kataoka2017} Kataoka, A., Tsukagoshi, T., Pohl, A., et al.\ 2017, \apjl, 844, L5 

%\bibitem[Kataoka et al.(2016a)]{Kataoka2016a} Kataoka, A., Muto, T., Momose, M., Tsukagoshi, T., \& Dullemond, C.~P.\ 2016a, \apj, 820, 54 

%\bibitem[Kataoka et al.(2016b)]{Kataoka2016b} Kataoka, A., Tsukagoshi, T., Momose, M., et al.\ 2016b, \apjl, 831, L12 

%\bibitem[Kataoka et al.(2015)]{Kataoka2015} Kataoka, A., Muto, T., Momose, M., et al.\ 2015, \apj, 809, 78 

\bibitem[Kataoka et al.(2012)]{Kataoka2012} Kataoka, A., Machida, M.~N., \& Tomisaka, K.\ 2012, \apj, 761, 40 





%\bibitem[Keto et al.(1988)]{Keto1988} Keto, E.~R., Ho, P.~T.~P., 
%\& Haschick, A.~D.\ 1988, \apj, 324, 920 
%\bibitem[Keto \& Zhang(2010)]{Keto2010} Keto, E., \& Zhang, Q.\ 2010, \mnras, 406, 102 

%\bibitem[Kounkel et al.(2017)]{Kounkel2017} Kounkel, M., Hartmann, L.,
%Loinard, L., et al.\ 2017, \apj, 834, 142



%\bibitem[Krasnopolsky \& K\"{o}nigl(2002)]{Krasnopolsky2002} Krasnopolsky, R., \& K\"{o}nigl, A.\ 2002, \apj, 580, 987

%\bibitem[Kumar, Anandarao \& Davis(1999)]{Kumar1999} Kumar, M. S. 
% N. , Anandarao, B. G. \& Davis, C. J. 1999, \aap, 344, L9 

\bibitem[Kwon et al.(2009)]{Kwon2009} Kwon, W., Looney, L.~W., Mundy, L.~G., Chiang, H.-F., \& Kemball, A.~J.\ 2009, \apj, 696, 841 




\bibitem[Kwon et al.(2019)]{Kwon2019} Kwon, W., Stephens, I., Tobin, J., et
al.\ 2019, submitted.




%\bibitem[Lada(1985)]{Lada1985} Lada, C. J. 1985, \araa, 23, 267 
%\bibitem[Lada \& Fich(1996)]{Lada1996} Lada, C. J. \& Fich, M.  
%  1996, \apj, 459, 638 

%\bibitem[Konigl \& Pudritz(2000)]{Konigl2000} Konigl, A., \& Pudritz, R.~E.\
%2000, Protostars and Planets IV, 759

%\bibitem[Launhardt et al.(2009)]{Launhardt2009} Launhardt, R., Pavlyuchenkov, Y., Gueth, F., et al.\ 2009, \aap, 494, 147 

\bibitem[Lee(2010)]{Lee2010HH111} Lee, C.-F.\ 2010, \apj, 725, 712 
%\bibitem[Lee et al.(2010)]{Lee2010HH211} Lee, C.-F., Hasegawa, T.~I., Hirano, N., et al.\ 2010, \apj, 713, 731 
\bibitem[Lee(2011)]{Lee2011} Lee, C.-F.\ 2011, \apj, 741, 62 
%\bibitem[Lee et al.(2000)]{Lee2000} Lee, C.-F., Mundy, L.G., Reipurth, B.,  Ostriker, E.C., \& Stone, J.M. 2000, \apj, 542, 925
%\bibitem[Lee et al.(2015)]{Lee2015} Lee, C.-F., Hirano, N., Zhang, Q., et al.\ 2015, \apj, 805, 186
%\bibitem[Lee et al.(2001)]{Lee2001} Lee, C.-F., Stone, J.~M., Ostriker, E.~C., \& Mundy, L.~G.\ 2001, \apj, 557, 429
%\bibitem[Lee et al.(2006)]{Lee2006} Lee, C.-F., Ho, P.~T.~P., 
%Beuther, H., Bourke, T.~L., Zhang, Q., Hirano, N., \& Shang, H.\ 2006, \apj, 639, 292

\bibitem[Lee et al.(2007)]{Lee2007} Lee, C.-F., Ho, P.~T.~P., Palau, A.,
Hirano, N., Bourke, T.~L., Shang, H., \& Zhang, Q.\ 2007, \apj, 670, 1188

%\bibitem[Lee et al.(2002)]{Lee2002} Lee, 
% C.-F., Mundy, L.~G., Stone, J.~M., \& Ostriker, E.~C.\ 2002, \apj, 576, 294 

%\bibitem[Lee et al.(2009)]{Lee2009HH111} Lee, C.-F., Mao, Y.-Y., \& Reipurth, B.\ 2009, \apj, 694, 1395 


\bibitem[Lee et al.(2009)]{Lee2009HH211} Lee, C.-F., Hirano, N., Palau, A., et al.\ 2009, \apj, 699, 1584 

%\bibitem[Lee et al.(2015)]{Lee2015Jet} Lee, C.-F., Hirano, N., Zhang, Q., et al.\ 2015, \apj, 805, 186 


\bibitem[Lee et al.(2014a)]{Lee2014HH212} Lee, C.-F., Hirano, N., Zhang, Q., et al.\ 2014a, \apj, 786, 114 


\bibitem[Lee et al.(2014b)]{Lee2014Mag} Lee, C.-F., Rao, R., Ching, T.-C., et al.\ 2014b, \apjl, 797, L9 
%\bibitem[Lee et al.(2008)]{Lee2008} Lee, C.-F., Ho, P.~T.~P., Bourke, T.~L., et al.\ 2008, \apj, 685, 1026 
%\bibitem[Lee et al.(2007)]{Lee2007} Lee, C.-F., Ho, P.~T.~P., Hirano, N., et al.\ 2007, \apj, 659, 499 
\bibitem[Lee et al.(2016)]{Lee2016} Lee, C.-F., Hwang, H.-C., \& Li, Z.-Y.\ 2016, \apj, 826, 213 

%\bibitem[Lee et al.(2017c)]{Lee2017Jet} Lee, C.-F., Ho, P.~T.~P., Li, Z.-Y., et al.\ 2017c, Nature Astronomy, 1, 0152 

\bibitem[Lee et al.(2017b)]{Lee2017COM} Lee, C.-F., Li, Z.-Y., Ho, P.~T.~P., et al.\ 2017b, \apj, 843, 27 

\bibitem[Lee et al.(2017a)]{Lee2017Disk} Lee, C.-F., Li, Z.-Y., Ho, P.~T.~P., et al.\ 2017a, Science Advances, 3, e1602935 

\bibitem[Lee et al.(2018a)]{Lee2018Bjet} Lee, C.-F., Hwang, H.-C., Ching, T.-C., et al.\ 2018a, Nature Communications, 9, 4636 

\bibitem[Lee et al.(2018b)]{Lee2018BDisk} Lee, C.-F., Li, Z.-Y., Ching, T.-C., Lai, S.-P., \& Yang, H.\ 2018b, \apj, 854, 56 

\bibitem[Lee et al.(2018c)]{Lee2018HH211Disk} Lee, C.-F., Li, Z.-Y., Hirano, N., et al.\ 2018c, \apj, 863, 94 

%\bibitem[Lee et al.(2018d)]{Lee2018Dwind} Lee, C.-F., Li, Z.-Y., Codella, C., et al.\ 2018, \apj, 856, 14 


%\bibitem[Leurini et al.(2016)]{Leurini2016} Leurini, S., Codella, C., Cabrit, S., et al.\ 2016, \aap, 595, L4 



%\bibitem[Lee, Myers, \& Tafalla(2001)]{LMT2001} Lee, C.~W., 
% Myers, P.~C., \& Tafalla, M.\ 2001, \apjs, 136, 703
%\bibitem[Lee et al.(2005)]{Lee2005a} Lee, C., Ho, P.~T.~P., \& 
% White, S.~M.\ 2005, \apj, 619, 948
%\bibitem[Lee \& Ho(2005)]{Lee2005b} Lee, C., \& Ho, P.~T.~P.,
%\ 2005, \apj, in press
%\bibitem[Lee \& Sahai(2003)]{Lee2003} Lee, C.-F.~\& Sahai, R.\ 
%2003, \apj, 586, 319 

%\bibitem[Levreault(1988)]{Levreault1988} Levreault, R. M. 1988, \apj, 
%  330, 897
%\bibitem[Li et al.(2013)]{Li2013} Li, Z.-Y., Krasnopolsky, R., 
%\& Shang, H.\ 2013, \apj, 774, 82 
%\bibitem[Lin et al.(1994)]{Lin1994} Lin, D.~N.~C., Hayashi, M., 
%Bell, K.~R., \& Ohashi, N.\ 1994, \apj, 435, 821
%\bibitem[Linke \& Goldsmith(1980)]{Linke1980} Linke, R.~A., \& 
% Goldsmith, P.~F.\ 1980, \apj, 235, 437  
%\bibitem[Li \& Shu(1996)]{Li1996} Li, Z. -Y.  \& Shu, F. H. 
%  1996, \apj, 472, 211 


\bibitem[Li et al.(2014)]{Li2014} Li, Z.-Y., Krasnopolsky, R., Shang, H., \& Zhao, B.\ 2014, \apj, 793, 130 




%\bibitem[L{\'o}pez-Sepulcre et al.(2015)]{Lopez-Sepulcre2015}
%L{\'o}pez-Sepulcre, A., Jaber, A.~A., Mendoza, E., et al.\ 2015, \mnras,
%449, 2438

\bibitem[Machida \& Matsumoto(2011)]{Machida2011} Machida, M.~N., \& Matsumoto, T.\ 2011, \mnras, 413, 2767 



%\bibitem[Majumdar et al.(2016)]{Majumdar2016} Majumdar, L., Gratier, P.,
%Vidal, T., et al.\ 2016, \mnras, 458, 1859


%\bibitem[Mardones et al.(1997)]{Mardones1997} Mardones, D., Myers, 
%P.~C., Tafalla, M., Wilner, D.~J., Bachiller, R., \& Garay, G.\ 1997, \apj, 
%489, 719
%\bibitem[Marvel et al.(in press)]{Marvelpress}Marvel, K.B., McCaughrean, M.J., 
%  \& Sargent, A.I., manuscript in preparation
%\bibitem[Masson \& Chernin(1993)]{Masson1993} Masson, C. R. \& 
%  Chernin, L. M. 1993, \apj, 414, 230 

\bibitem[Masson et al.(2016)]{Masson2016} Masson, J., Chabrier, G., Hennebelle, P., Vaytet, N., \& Commer{\c c}on, B.\ 2016, \aap, 587, A32 


%\bibitem[Mathieu et al.(1988)]{Mathieu1988} Mathieu, R. D., Myers, 
%  P. C., Schild, R. E., Benson, P. J. \& Fuller, G. A. 1988, \apj, 330, 385 

%\bibitem[Matthews et al.(2010)]{Matthews2010} Matthews, L.~D., Greenhill,
%L.~J., Goddi, C., et al.\ 2010, \apj, 708, 80

\bibitem[Matthews et al.(2009)]{Matthews2009} Matthews, B.~C., McPhee, C.~A., Fissel, L.~M., \& Curran, R.~L.\ 2009, \apjs, 182, 143 




%\bibitem[Matzner \& McKee(1999)]{Matzner1999} Matzner, C. D. \& 
%  McKee, C. F. 1999, \apjl, 526, L109 
%\bibitem[McCaughrean et al.(2002)]{McCaughrean2002} McCaughrean, M., 
%Zinnecker, H., Andersen, M., Meeus, G., \& Lodieu, N.\ 2002, The Messenger, 109, 28

%\bibitem[McCaughrean et al.(1994)]{McCaughrean1994} McCaughrean, M.~J.,
%Rayner, J.~T., \& Zinnecker, H.\ 1994, \apjl, 436, L189




\bibitem[Maury et al.(2018)]{Maury2018} Maury, A.~J., Girart, J.~M., Zhang,
Q., et al.\ 2018, \mnras, 477, 2760




%\bibitem[McKee et al.(1987)]{Mckee1987} McKee, 
%  C. F., Hollenbach, D. J., Seab, G. C. \& Tielens, A. G. G. M. 1987, \apj, 
%  318, 674
\bibitem[Mellon \& Li(2008)]{Mellon2008} Mellon, R.~R., \& Li, Z.-Y.\ 2008, \apj,
681, 1356 

%\bibitem[Mendoza et al.(2014)]{Mendoza2014} Mendoza, E., Lefloch, B., L{\'o}pez-Sepulcre, A., et al.\ 2014, \mnras, 445, 151 

%\bibitem[Menten et al.(2007)]{Menten2007} Menten, K.~M., Reid, M.~J., Forbrich, J., \&
%Brunthaler, A.\ 2007, \aap, 474, 515 
%\bibitem[Meyers-Rice \& Lada(1991)]{Meyers1991} Meyers-Rice, B. A. 
%\& Lada, C. J. 1991, \apj, 368, 445 
%\bibitem[Micono et al.(1998)]{Micono1998} Micono, M., Davis, C. J., 
%Ray, T. P., Eisloeffel, J. \& Shetrone, M. D. 1998, \apjl, 494, L227 
%\bibitem[Momose et al.(1998)]{Momose1998} Momose, M., Ohashi, N., 
% Kawabe, R., Nakano, T., \& Hayashi, M.\ 1998, \apj, 504, 314 

%\bibitem[M{\"u}ller et al.(2016)]{Muller2016} M{\"u}ller, H.~S.~P.,
%Belloche, A., Xu, L.-H., et al.\ 2016, \aap, 587, A92


%\bibitem[Murillo et al.(2013)]{Murillo2013} Murillo, N.~M., Lai, S.-P., Bruderer, S., Harsono, D., \& van Dishoeck, E.~F.\ 2013, \aap, 560, A103 
%\bibitem[Murillo \& Lai(2013)]{Murillo2013a} Murillo, N.~M., \& Lai, S.-P.\ 2013, \apjl, 764, L15 
%\bibitem[Myers et al.(1987)]{Myers1987} Myers, P.~C., Fuller, 
%G.~A., Mathieu, R.~D., Beichman, C.~A., Benson, P.~J., Schild, R.~E., \& 
%Emerson, J.~P.\ 1987, \apj, 319, 340 
%\bibitem[Myers et al.(1988)]{Myers1988} Myers, 
%\bibitem[Myers \& Ladd(1993)]{Myers1993} Myers, P.~C.~\& Ladd, 
%E.~F.\ 1993, \apjl, 413, L47 
%P. C., Heyer, M., Snell, R. L. \& Goldsmith, P. F. 1988, \apj, 324, 907 
%\bibitem[Myers et al.(2000)]{Myers2000} Myers, P.~C., Evans, 
%N.~J., \& Ohashi, N.\ 2000, Protostars and Planets IV, 217 
%\bibitem[Nagar et al.(1997)]{Nagar1997} Nagar, N. 
% M., Vogel, S. N., Stone, J. M. \& Ostriker, E. C. 1997, \apjl, 482, L195 
%\bibitem[Ostriker(1997)]{Ostriker1997} Ostriker, E. C. 1997, \apj, 486, 291

%\bibitem[Nakamura(2000)]{Nakamura2000} Nakamura, F.\ 2000, \apj, 543, 291

\bibitem[Ohashi et al.(1997)]{Ohashi1997} Ohashi, N., Hayashi, M., Ho, P.~T.~P., \& Momose, M.\ 1997, \apj, 475, 211 

\bibitem[Ohashi et al.(2014)]{Ohashi2014} Ohashi, N., Saigo, K., Aso, Y., et
al.\ 2014, \apj, 796, 131

%\bibitem[Ostriker et. al.(2001)]{Ostriker2001} 
% Ostriker, E.~C., Lee, C., Stone, J.~M., \& Mundy, L.~G.\ 2001, \apj, 557, 
% 443 

\bibitem[Ortiz-Le{\'o}n et al.(2018)]{Ortiz-Leon2018} Ortiz-Le{\'o}n, G.~N.,
Loinard, L., Dzib, S.~A., et al.\ 2018, \apj, 865, 73

\bibitem[Ossenkopf \& Henning(1994)]{Ossenkopf1994} Ossenkopf, V., \& Henning, T.\ 1994, \aap, 291, 943 


%\bibitem[Oya et al.(2016)]{Oya2016} Oya, Y., Sakai, N., L{\'o}pez-Sepulcre,
%A., et al.\ 2016, \apj, 824, 88



%\bibitem[Palau et al.(2006)]{Palau2006} Palau, A., et al.\ 2006, \apjl, 636, L137

%\bibitem[Panoglou et al.(2012)]{Panoglou2012} Panoglou, D., Cabrit, S.,
%Pineau Des For{\^e}ts, G., et al.\ 2012, \aap, 538, A2




%\bibitem[Parise et al.(2006)]{Parise2006} Parise, B., Ceccarelli, C., Tielens, A.~G.~G.~M., et al.\ 2006, \aap, 453, 949 

%\bibitem[Pineau des For$\hat {\rm e}$ts et al.(1993)]{Pineau1993} Pineau des For$\hat {\rm e}$ts G., Roueff E., Schilke P., \& Flower D.R. 1993, MNRAS 262, 915

%\bibitem[Podio et al.(2015)]{Podio2015} Podio, L., Codella, C., Gueth, F.,
%et al.\ 2015, \aap, 581, A85


%\bibitem[Pohl et al.(2016)]{Pohl2016} Pohl, A., Kataoka, A., Pinilla, P., et al.\ 2016, \aap, 593, A12 


%\bibitem[Pudritz et al.(2007)]{Pudritz2007} Pudritz, R.~E., Ouyed, 
%R., Fendt, C., \& Brandenburg, A.\ 2007, Protostars and Planets V, 277 
%\bibitem[Pirogov et al.(2003)]{Pirogov2003} Pirogov, L., Zinchenko, 
%I., Caselli, P., Johansson, L.~E.~B., \& Myers, P.~C.\ 2003, \aap, 405, 639 
%\bibitem[Raga et al.(1990)]{Raga1990} Raga, A.~C., Binette, L., 
%Canto, J., \& Calvet, N.\ 1990, \apj, 364, 601
%\bibitem[Raga (1993)]{Raga1993A} Raga, A. C. 1993, \apss, 208, 163 
%\bibitem[Raga \& Cabrit(1993)]{Raga1993} Raga, A. \& Cabrit, S. 
%1993, \aap, 278, 267 
%\bibitem[Raga et al.(1993b)]{Raga1993b} Raga, A. C., Canto, J., Calvet N.,
%Rodriguez, L.F., \& Torrelles, J.M., 1993b, \aap, 276, 539 
%\bibitem[Raga et al.(2004)]{Raga2004} Raga, A.~C., Noriega-Crespo, A., 
%Gonz{\' a}lez, R.~F., \& Vel{\' a}zquez, P.~F.\ 2004, \apjs, 154, 346 
%\bibitem[Raga, Cabrit \& Canto(1995)]{Raga1995} Raga, A. C., 
%Cabrit, S. \& Canto, J. 1995, \mnras, 273, 422


%\bibitem[Rao et al.(2014)]{Rao2014} Rao, R., Girart, J.~M., Lai, S.-P., \& Marrone, D.~P.\ 2014, \apjl, 780, L6 

%\bibitem[Ray et al.(2007)]{Ray2007} Ray, T., Dougados, C., 
%Bacciotti, F., Eisl{\"o}ffel, J., 
%\& Chrysostomou, A.\ 2007, Protostars and Planets V, 231 
%\bibitem[Rawlings, Taylor, \& Williams(2000)]{Rawlings2000} 
%Rawlings, J.~M.~C., Taylor, S.~D., \& Williams, D.~A.\ 2000, \mnras, 313, 
%461
%\bibitem[Rawlings et al.(2004)]{Rawlings2004} Rawlings, J.~M.~C., 
%Redman, M.~P., Keto, E., \& Williams, D.~A.\ 2004, \mnras, 351, 1054 
%\bibitem[Rawlings et al.(2013)]{Rawlings2013} Rawlings, J.~M.~C., 
%Redman, M.~P., \& Carolan, P.~B.\ 2013, \mnras, 435, 289 

%\bibitem[Rebull et al.(2007)]{Rebull2007} Rebull, L.~M., Stapelfeldt, K.~R., Evans, N.~J., II, et al.\ 2007, \apjs, 171, 447 




%\bibitem[Reipurth \& Olberg(1991)]{Reipurth1991} Reipurth, B. \& 
%Olberg, M.  1991, \aap, 246, 535 
%\bibitem[Reipurth \& Cernicharo(1995)]{Reipurth1995} Reipurth, B. \& 
% Cernicharo, J. 1995, Revista Mexicana de Astronomia y Astrofisica 
% Conference Series, 1, 43  
%\bibitem[Reipurth, Raga, \& Heathcote(1992)]{Reipurth1992} Reipurth, 
%  B., Raga, A. C. \& Heathcote, S.  1992, \apj, 392, 145 
%\bibitem[Reipurth Bally \& Devine(1997a)]{Reipurth1997a} Reipurth, B., 
%Bally, J.  \& Devine, D.  1997a, \aj, 114, 2708 
%\bibitem[Reipurth et al.(1993)]{Reipurth1993}Reipurth, B., Chini, R., Krugel, E., 
%  Kreysa, E., Sievers., A. 1993 A\&A, 273, 221
%\bibitem[Reipurth et al.(1997b)]{Reipurth1997b} Reipurth, B., Hartigan, 
%  P. , Heathcote, S. , Morse, J. A. \& Bally, J.  1997b, \aj, 114, 757 
%\bibitem[Reipurth et al.(1999)]{Reipurth1999} Reipurth, B., Yu, K. , 
%Rodriguez, L. F., Heathcote, S.  \& Bally, J.  1999, \aap, 352, L83 
%\bibitem[Reipurth et al.(2002)]{Reipurth2002} Reipurth, B., 
%Heathcote, S., Morse, J., Hartigan, P., \& Bally, J.\ 2002, \aj, 123, 362 
%\bibitem[Richer et al.(2000)]{Richer2000}Richer, J. Shepherd, D., 
%  Cabrit, S., Bachiller, R., \& Churchwell, E. 2000, in Protostars
%and Planets IV, ed. V. Mannings, A. P. Boss \& S. S. Russell
%  (Tucson: University of Arizona Press), in press 
%\bibitem[Rodriguez \& Reipurth(1994)]{Rodriguez1994} Rodriguez, L. F. 
% \& Reipurth, B. 1994, \aap, 281, 882 
%\bibitem[Rodriguez et al.(1998)]{Rodriguez1998} Rodriguez, L. F., 
% Reipurth, B., Raga, A. C. \& Canto, J. 1998, Revista Mexicana de Astronomia 
% y Astrofisica, 34, 69 
%\bibitem[Rodriguez \& Reipurth(1998)]{Rodriguez1998} Rodriguez, L. F., 
% \& Reipurth, B. 1998, Revista Mexicana de Astronomia 
% y Astrofisica, 34, 13 


\bibitem[Sadavoy et al.(2018)]{Sadavoy2018} Sadavoy, S.~I., Myers, P.~C., Stephens, I.~W., et al.\ 2018, \apj, 859, 165 

%\bibitem[Sadavoy et al.(2018)]{2018ApJ...869..115S} Sadavoy, S.~I., Myers, P.~C., Stephens, I.~W., et al.\ 2018, \apj, 869, 115 

\bibitem[Sakai et al.(2014)]{Sakai2014} Sakai, N., Sakai, T., Hirota, T., et al.\ 2014, \nat, 507, 78
%\bibitem[Sakai et al.(2014b)]{Sakai2014b} Sakai N., Oya Y., Sakai T., et al. 2014b, ApJ 791, L38
%\bibitem[Sakai et al.(2016)]{Sakai2016} Sakai N., Oya Y., L\'opez-Sepulcre A., et al. 2016, MNRAS 820, L34
%\bibitem[Sakai et al.(2017)]{Sakai2017} Sakai, N., Oya, Y., Higuchi, A.~E., et al.\ 2017, \mnras, 467, L76

%\bibitem[Saladino(2012)]{Saladino2012} Saladino R.  , Botta G.  , Pino S.  ,
%Costanzo G.  , Di Mauro E..  , Chem Soc Rev , 2012, vol.  41 pg.  5526

%Saladino, R.\ 2012, 39th COSPAR Scientific Assembly, 39, 1657 

%\bibitem[Sandford \& Allamandola(1993)]{Sandford1993} Sandford, S.~A., \&
%Allamandola, L.~J.\ 1993, \apj, 417, 815

%\bibitem[Sandstrom et al.(2007)]{Sandstrom2007} Sandstrom, K.~M., 
%Peek, J.~E.~G., Bower, G.~C., Bolatto, A.~D., 
%\& Plambeck, R.~L.\ 2007, \apj, 667, 1161 
%\bibitem[Schilke et al.(1997)]{Schilke1997} Schilke, P., Walmsley, 
%C.~M., Pineau des Forets, G., \& Flower, D.~R.\ 1997, \aap, 321, 293 


%\bibitem[Segura-Cox et al.(2015)]{Segura2015} Segura-Cox, D.~M., Looney, L.~W., Stephens, I.~W., et al.\ 2015, \apjl, 798, L2 

\bibitem[Segura-Cox et al.(2018)]{Segura-Cox2018} Segura-Cox, D.~M., Looney, L.~W., Tobin, J.~J., et al.\ 2018, \apj, 866, 161 


%\bibitem[Sewi{\l}o et al.(2017)]{Sewilo2017} Sewi{\l}o, M., Wiseman, J., Indebetouw, R., et al.\ 2017, \apj, 849, 68 



%\bibitem[Shang et al.(1998)]{Shang1998} Shang, H. , Shu, 
%  F. H. \& Glassgold, A. E. 1998, \apjl, 493, L91 
%\bibitem[Shang et al.(2004)]{Shang2004} Shang, H., Lizano, S., 
% Glassgold, A., \& Shu, F.\ 2004, \apjl, 612, L69 
%\bibitem[Shang et al.(2006)]{Shang2006} Shang, H., Allen, A., Li, 
%Z.-Y., Liu, C.-F., Chou, M.-Y., \& Anderson, J.\ 2006, \apj, 649, 845 
%\bibitem[Shang et al.(2007)]{Shang2007} Shang, H., Li, Z.-Y., 
%\& Hirano, N.\ 2007, Protostars and Planets V, 261 
%\bibitem[Shirley et al.(2000)]{Shirley2000} 
% Shirley, Y.~L., Evans, N.~J., Rawlings, J.~M.~C., \& Gregersen, E.~M.\ 
% 2000, ApJS, 131, 249 

\bibitem[Shu(1977)]{Shu1977} Shu, F.~H.\ 1977, \apj, 214, 488 

%\bibitem[Shu, Adams, \& Lizano(1987)]{Shu1987} Shu, F.~H., 
% Adams, F.~C., \& Lizano, S.\ 1987, \araa, 25, 23 
%\bibitem[Shu et al.(1991)]{Shu1991} Shu, F. H., 
%  Ruden, S. P., Lada, C. J. \& Lizano, S.  1991, \apjl, 370, L31 
%\bibitem[Shu et al.(1994)]{Shu1994} Shu, F., Najita, J., 
%Ostriker, E., et al.\ 1994, \apj, 429, 781 
%\bibitem[Shu et al.(1995) 1995]{Shu1995} Shu, F. 
%  H., Najita, J. , Ostriker, E. C. \& Shang, H.  1995, \apjl, 455, L155  
%\bibitem[Shu et al.(2000)]{Shu2000} Shu, F.~H., Najita, J.~R., 
%Shang, H., \& Li, Z.-Y.\ 2000, Protostars and Planets IV, 789 
%\bibitem[Simon et al.(2000)]{Simon2000} Simon, M., Dutrey, A., 
% \& Guilloteau, S.\ 2000, \apj, 545, 1034 

%\bibitem[Simon et al.(2013a)]{Simon2013a}
%Simon, J. B., Bai, X.-N., Armitage, P. J. et al. 2013, ApJ, 764, 66

%\bibitem[Simon et al.(2013b)]{Simon2013b}
%Simon, J. B., Bai, X.-N., Armitage, P. J. et al. 2013, ApJ, 775, 73


%\bibitem[Smith, Suttner, \& Yorke(1997)]{Smith1997} Smith, M. D., 
%  Suttner, G. \& Yorke, H. W. 1997, \aap, 323, 223 
%\bibitem[Smith \& Rosen(2007)]{Smith2007} Smith, M.~D., \& Rosen, 
%A.\ 2007, \mnras, 378, 691 

%\bibitem[Stephens et al.(2014)]{Stephens2014} Stephens, I.~W., Looney,
%L.~W., Kwon, W., et al.\ 2014, \nat, 514, 597



\bibitem[Stephens et al.(2017)]{Stephens2017} Stephens, I.~W., Yang, H., Li, Z.-Y., et al.\ 2017, \apj, 851, 55 





%\bibitem[Steer, Dewdney, \& Ito(1984)]{Steer1984} Steer, D.~G., 
% Dewdney, P.~E., \& Ito, M.~R.\ 1984, \aap, 137, 159 
%\bibitem[Stone \& Norman(1994)]{Stone1994} Stone, J. M. \& Norman, 
%M. L. 1994, \apj, 420, 237 

%\bibitem[Suttner et al.(1997)]{Suttner1997}  Suttner, G., Smith, M. D.,
%  Yorke, H.  W.  \& Zinnecker, H.  1997, \aap, 318, 595


%\bibitem[Tabone et al.(2017)]{Tabone2017} Tabone, B., Cabrit, S., Bianchi, E., et al., 
%2017, \aap, arXiv:1710.01401

%\bibitem[Tafalla et al.(2002)]{Tafalla2002} Tafalla, M., Myers, 
%P.~C., Caselli, P., Walmsley, C.~M., \& Comito, C.\ 2002, \apj, 569, 815
%\bibitem[Takakuwa et al.(2003)]{Takakuwa2003} Takakuwa, S., 
%Kamazaki, T., Saito, M., \& Hirano, N.\ 2003, \apj, 584, 818
%\bibitem[Takami et al.(2006)]{Takami2006} Takami, M., Takakuwa, 
%S., Momose, M., Hayashi, M., Davis, C., Pyo, T.-S., Nishikawa, T., \& 
%Kohno, K.\ 2006, PASJ, ??





%\bibitem[Taylor 
%\& Raga(1995)]{Taylor1995} Taylor, S.~D., \& Raga, A.~C.\ 1995, \aap, 296, 823 

%\bibitem[Tazaki et al.(2017)]{Tazaki2017} Tazaki, R., Lazarian, A., \&
%Nomura, H.\ 2017, \apj, 839, 56


%\bibitem[Terebey et al.(1984)]{Terebey1984} Terebey, S., Shu, F.~H., \& Cassen, P.\ 1984, \apj, 286, 529 
%\bibitem[Terebey et al.(1993)]{Terebey1993} Terebey, S.,  Chandler, C. J. \& Andre, P. 1993, \apj, 414, 759 

%\bibitem[Tieftrunk et al.(2006)]{Tieftrunk1994} Tieftrunk A., Pineau des For$\hat {\rm e}$ts G., Schilke P., \& Walmsley C.M. 1994, A\&A 289, 579

\bibitem[Tobin et al.(2012)]{Tobin2012} Tobin, J.~J., Hartmann, L., Chiang, H.-F., et al.\ 2012, \nat, 492, 83 

\bibitem[Tobin et al.(2015)]{Tobin2015} Tobin, J.~J., Looney, L.~W., Wilner, D.~J., et al.\ 2015, \apj, 805, 125 





%\bibitem[Todo et al.(1993)]{Todo1993} Todo, Y., Uchida, Y., 
%Sato, T., \& Rosner, R.\ 1993, \apj, 403, 164 


\bibitem[Tomisaka(2011)]{Tomisaka2011} Tomisaka, K.\ 2011, \pasj, 63, 147 


\bibitem[Tanner \& Arce(2011)]{Tanner2011} Tanner, J.~D., \& Arce, H.~G.\ 2011, \apj, 726, 40 


%\bibitem[Ulrich(1976)]{Ulrich1976} Ulrich, R.~K.\ 1976, \apj, 210, 377 
%\bibitem[Umemoto et al.(1991)]{Umemoto1991} Umemoto, T., Hirano, 
%  N., Kameya, O., Fukui, Y., Kuno, N. \& Takakubo, K. 1991, \apj, 377, 510

%\bibitem[van Dishoeck \& Blake(1998)]{vanDishoeck1998} 
%van Dishoeck E.F., \& Blake G.A. 1998, ARA\&A 36, 317


\bibitem[V{\"a}is{\"a}l{\"a} et al.(2019)]{Vaisala2019} V{\"a}is{\"a}l{\"a}, M.~S., Shang, H., Krasnopolsky, R., et al.\ 2019, \apj, 873, 114 


%\bibitem[Velusamy \& Langer(1998)]{Velusamy1998} Velusamy, T. \& 
%  Langer, W. D. 1998, \nat, 392, 685 
%\bibitem[Viti, Natarajan, \& Williams(2002)]{Viti2002} Viti, S., 
% Natarajan, S., \& Williams, D.~A.\ 2002, \mnras, 336, 797 
%\bibitem[\Volker{} et al.(1999)]{Volker1999}
%  \Volker, R. , Smith, M. D., Suttner, G.  \& Yorke, H. W. 1999, 
%  \aap, 343, 953 

%\bibitem[Wakelam et al.(2004)]{Wakelam2004}
%Wakelam V., Caselli P., Ceccarelli C., Herbst E., \& Castets A. 2004, A\&A 422, 159

%\bibitem[Walsh et al.(2016)]{Walsh2016} Walsh, C., Loomis, R.~A., {\"O}berg, K.~I., et al.\ 2016, \apjl, 823, L10 

%\bibitem[Ward-Thompson, Motte, \& Andre(1999)]{Ward-Thompson1999} 
% Ward-Thompson, D., Motte, F., \& Andre, P.\ 1999, \mnras, 305, 143 
%\bibitem[Weintraub et al.(1994)]{Weintraub1994} 
%  Weintraub, D. A., Tegler, S. C., Kastner, J. H. \& Rettig, T.  1994, \apj, 
%423, 674 
\bibitem[Wiseman et al.(2001)]{Wiseman2001} Wiseman, J., Wootten, 
 A., Zinnecker, H., \& McCaughrean, M.\ 2001, \apjl, 550, L87 
%\bibitem[Wootten et al.(2002)]{Wootten2002} 
% Wootten, A., Mangum, J.~G., Wiseman, J., \& Fuller, G.~A.\ 2002, Bulletin 
% of the American Astronomical Society, 34, 1134 
%\bibitem[Wouterloot \& Walmsley(1986)]{Wouterloot1986} Wouterloot, J. 
%  G. A. \& Walmsley, C. M. 1986, \aap, 168, 237 
%\bibitem[Wu, Huang \& He(1996)]{Wu1996} Wu, Y., Huang, M. \& He, 
%  J. 1996, \aaps, 115, 283 
%\bibitem[Xu et al.(2000)]{Xu2000} Xu, J., Hardee, P.~E., \& 
%Stone, J.~M.\ 2000, \apj, 543, 161 

%\bibitem[Xu et al.(2012)]{Xu2012} Xu, L.-H., Lees, R.~E., Crabbe, G.T.,
%Myshrall, J.A., et al.  \ 2012, {\it J.  Chem.  Phys}, 137, 104313


%\bibitem[Yang et al.(1997)]{Yang1997} Yang, J., Ohashi, N. , Yan, 
% J. , Liu, C. , Kaifu, N.  \& Kimura, H.  1997, \apj, 475, 683 


%\bibitem[Yang et al.(2016a)]{Yang2016a} Yang, H., Li, Z.-Y., Looney, L.~W.,
%et al.\ 2016a, \mnras, 460, 4109


%\bibitem[Yang et al.(2016b)]{Yang2016b} Yang, H., Li, Z.-Y., Looney, L., \&
%Stephens, I.\ 2016b, \mnras, 456, 2794

%\bibitem[Yang et al.(2017)]{Yang2017} Yang, H., Li, Z.-Y., Looney, L.~W.,
%Girart, J.~M., \& Stephens, I.~W.\ 2017, \mnras, 472, 373



%\bibitem[Yen et al.(2017)]{Yen2017} Yen, H.-W., Koch, P.~M., Takakuwa, S.,
%et al.\ 2017, \apj, 834, 178

\bibitem[Yen et al.(2015)]{Yen2015} Yen, H.-W., Takakuwa, S., Koch, P.~M.,
et al.\ 2015, \apj, 812, 129




%\bibitem[Zinnecker, McCaughrean, \& Rayner(1998)]{Zinnecker1998} 
%  Zinnecker, H., McCaughrean, M. J. \& Rayner, J. T. 1998, \nat, 394, 862 
%\bibitem[Zinnecker et al.(1992)]{Zinnecker1992} Zinnecker, H. , 
%  Bastien, P. , Arcoragi, J. -P.  \& Yorke, H. W. 1992, \aap, 265, 726 
\end{thebibliography}
\end{document}